\documentclass[12pt,a4paper]{article}
\pdfoutput=1

\usepackage{geometry}
\usepackage{amsmath}
\usepackage{amssymb}
\usepackage[dvips]{graphicx}
\usepackage{cite}
\usepackage{pstricks}
\usepackage{bm}
\usepackage{pbox}
\usepackage{placeins}
\usepackage{graphicx}
\usepackage{caption}
\usepackage{subcaption}
\usepackage[T1]{fontenc}
\usepackage{footnote}
\usepackage{pdfpages}
\usepackage{hhline}
\usepackage{multirow}
\usepackage{enumitem}
\usepackage[toc,page]{appendix}
\allowdisplaybreaks

\usepackage[perpage]{footmisc}

\definecolor{MyDarkBlue}{rgb}{0.1, 0.1, 0.8} 
\definecolor{MyLightBlue}{rgb}{0.22,0.51,0.9}
\usepackage[colorlinks=true,linkcolor=blue,citecolor=MyDarkBlue,
urlcolor=MyLightBlue,bookmarksnumbered=true,bookmarksopen]{hyperref}
\hypersetup{colorlinks, citecolor=red,linkcolor=MyDarkBlue, urlcolor=MyLightBlue}
\usepackage{wasysym}

\def\Tr{\mbox{Tr}\,}

\newcommand\lsim{\mathrel{\rlap{\lower4pt\hbox{\hskip1pt$\sim$}}
    \raise1pt\hbox{$<$}}}
\newcommand\gsim{\mathrel{\rlap{\lower4pt\hbox{\hskip1pt$\sim$}}
    \raise1pt\hbox{$>$}}}

\newcommand{\beq}{\begin{equation}}
\newcommand{\eeq}{\end{equation}}
\newcommand{\bet}{\begin{itemize}}
\newcommand{\eet}{\end{itemize}}
\newcommand{\ben}{\begin{enumerate}}
\newcommand{\een}{\end{enumerate}}
\newcommand{\bea}{\begin{eqnarray}}
\newcommand{\eea}{\end{eqnarray}}
\newcommand{\bem}{\begin{pmatrix}}
\newcommand{\eem}{\end{pmatrix}}
\newcommand{\noi}{\noindent}
\newcommand{\non}{\nonumber}

\newcommand\scalemath[2]{\scalebox{#1}{\mbox{\ensuremath{\displaystyle #2}}}}

\begin{document}

\numberwithin{equation}{section}

\vspace*{-0.2in}
\begin{flushright}
OSU-HEP-18-03
\end{flushright}
\vspace{0.5cm}
\begin{center}

{\Large\bf Resurrecting Minimal Yukawa Sector of SUSY SO(10)}
\vspace{1cm}

\renewcommand{\thefootnote}{\fnsymbol{footnote}}
\centerline{
{}~K.S. Babu$^{a,}$\footnote{E-mail: \textcolor{MyLightBlue}{babu@okstate.edu}},{}~
Borut Bajc$^{b,}$\footnote{E-mail: \textcolor{MyLightBlue}{borut.bajc@ijs.si}} and
{}~Shaikh Saad$^{a,}$\footnote{E-mail: \textcolor{MyLightBlue}{shaikh.saad@okstate.edu}}
}

\vspace{0.5cm}
\centerline{$^{a}${\it\small Department of Physics, Oklahoma State University, Stillwater, OK, 74078, USA }}
\centerline{$^{b}${\it\small Jo\v{z}ef Stefan Institute, 1000 Ljubljana, Slovenia}}
 \end{center}

\renewcommand{\thefootnote}{\arabic{footnote}}

\bigskip

\begin{abstract}
{\footnotesize
Supersymmetric $SO(10)$ models with Yukawa coupling matrices involving only a $10_H$ and a $\overline{126}_H$ of Higgs fields
can lead to a predictive and consistent scenario for fermion masses and mixings, including the neutrino sector.  However, when coupled minimally
to a symmetry breaking sector that includes a $210_H$ and a $126_H$, these models lead either to an unacceptably small neutrino mass scale, or to non-perturbative values of the gauge couplings.  Here we show that with the addition of a $54_H$ to the symmetry breaking sector, the successful predictions of these models for fermion masses and mixings can be maintained.  The $54_H$ enables a reduction of the $B-L$ symmetry breaking scale to an intermediate value of order $10^{12}$ GeV, consistent with the observed neutrino mass spectrum, while preserving perturbative gauge coupling unification. We obtain an excellent fit to all fermion masses and mixings in this framework. We analyze carefully the prediction of the model for CP violation in neutrino oscillations. Consistency with proton lifetime, however, requires a mini-split SUSY spectrum with the squarks and sleptons having masses of order 100 TeV, accompanied by TeV scale gauginos and Higgsinos.  Such a spectrum may arise from pure gravity mediation, which would predict the partial lifetime for the decay $p \rightarrow \overline{\nu} K^+$ to be an order of magnitude above the current experimental limit.
}
 \end{abstract}

\clearpage


\section{Introduction}

Perhaps the most attractive feature of unified models based on $SO(10)$ gauge symmetry is that all fermions of a given family, including the right-handed neutrino, are assembled into a single representation, the ${\bf 16}$--dimensional spinor of $SO(10)$.  The Higgs fields that can generate masses for the fermions can be inferred from the fermion bilinears,
\begin{equation}
16 \times 16 = 10_s + 120_a + 126_s~,
\end{equation}
where the subscripts $s$ and $a$ denote symmetric and antisymmetric combinations.  At least two Higgs fields are needed to generate nontrivial quark and lepton mixings.  A minimal Yukawa sector in supersymmetric (SUSY) $SO(10)$, which is the focus of this paper, is thus obtained by the introduction of a $10_H$ and a $\overline{126}_H$ of Higgs fields.\footnote{Choosing any other combinations involving two of these Higgs fields will not lead to viable phenomenology. For example, using two copies of $10_H$ will lead to nonzero quark mixings, but the unviable mass relations $m_\mu^0 = m_s^0$ and $m_e^0 = m_d^0$ will prevail in this case at the GUT scale. If two copies of $\overline{126}_H$ is used instead, unacceptable relations $m_\tau^0 = -3 m_b^0$ and $m_e^0 = - 3 m_d^0$ will emerge. Similarly, combining a $120_H$ with either $10_H$ or $\overline{126}_H$ will also lead to inconsistent phenomenology.}  In SUSY $SO(10)$ both the $10_H$ and $\overline{126}_H$ are complex chiral superfields, resulting in the Yukawa superpotential:
\beq
W_{SO(10)}^{{\rm Yukawa}}=16^T\left(Y_{10}\,10_H+Y_{126}\overline{126}_H\,\right)16~.
\label{sup}
\eeq
Here $Y_{10}$ and $Y_{126}$ are $3 \times 3$ symmetric matrices in family space.  These two matrices will be responsible for generating all of the charged fermion masses and mixings as well as the heavy right-handed and light left-handed neutrino masses and mixings \cite{Babu:1992ia}.  The mass matrices arising from Eq. (\ref{sup}) are displayed in detail in Eqs. (\ref{y1})-(\ref{y2}).  This simple Yukawa structure has 12 real parameters and 7 phases, which should fit 18 measured quantities -- 6 quark masses, 4 quark mixing parameters, 3 charged lepton masses, 2 neutrino mass splittings and 3 leptonic mixing angles.  (One Yukawa coupling matrix, say $Y_{126}$, can be made diagonal and real, whence the other would have 6 complex parameters.  In addition to the Yukawa couplings, the fermion mass matrices will involve 3 ratios of vacuum expectation values (VEVs), with one ratio remaining complex.  This leads to a total of 12 real parameters and 7 phases.) It is remarkable that this minimal setup is able to reproduce the full fermion mass spectrum, including large leptonic mixing angles along with small quark mixing angles \cite{Bajc:2002iw,Fukuyama:2002ch,Goh:2003sy,Goh:2003hf,
Bertolini:2004eq,Babu:2005ia,Bertolini:2006pe,Joshipura:2011nn,Altarelli:2013aqa,Dueck:2013gca,Bajc:2008dc}, a nontrivial feat in any quark--lepton unified framework. To its credit, this setup predicted \cite{Goh:2003sy,Goh:2003hf,Babu:2005ia,Bertolini:2006pe,Joshipura:2011nn,Bajc:2008dc} a large value for the reactor neutrino mixing angle $\theta_{13}$ very close to the experimental value measured subsequently by the Daya Bay collaboration \cite{An:2012eh}.

A crucial feature of Eq. (\ref{sup}) in generating the mass matrices of Eqs. (\ref{y1})-(\ref{y2}) is that the Standard Model (SM) singlet in $\overline{126}_H$ acquires a large VEV ($V_R$) of order $10^{12}$ GeV, while its SM doublet components also acquire weak scale VEVs \cite{Babu:1992ia}. This is possible only when the light MSSM Higgs fields $H_u$ and $H_d$ emerge at least partially from the Higgs doublet fragments of the $\overline{126}_H$.  The simplest choice for generating such mixings among Higgs doublets of $10_H$ and $\overline{126}_H$ is via a renormalizable superpotential coupling involving the $210_H$ field, viz., $W \supset \lambda_4 \,\overline{126}_H \,210_H \,10_H$.  Thus, a minimal renormalizable symmetry breaking sector is identified: $\{10_H + \overline{126}_H + 126_H + 210_H\}$.  Here the $126_H$ is needed in order to break $SO(10)$ gauge symmetry in the SUSY limit. The $\{\overline{126}_H + 126_H + 210_H\}$ fields jointly break $SO(10)$ symmetry down to the SM, while the $SU(2)_L$ doublets from these fields along with the doublets from the $10_H$ break the electroweak symmetry at a lower scale. This model has been extensively studied in the literature, and has been referred to as the minimal SUSY $SO(10)$ model.

It has long been recognized \cite{Aulakh:1982sw,Clark:1982ai,Babu:1992ia} that this minimal SUSY $SO(10)$ model contains all the essential ingredients to be realistic.  Furthermore, it was shown in Ref. \cite{Aulakh:2003kg} that this model has the minimal number of parameters among all supersymmetric grand unified theories.\footnote{Here and in the following we will consider only renormalizable models. In fact, any non-renormalizable GUT will have innumerable parameters.} The supersymmetric sector of this model is described by 26 real parameters: $3+12 = 15$ real Yukawa couplings and $14-3 = 11$ real superpotential parameters from the symmetry breaking sector $W_{SSB}$.  There are 7 complex parameters in $W_{SSB}$, of which 3 can be made real. Significant effort has been put into the study of the parameter space of this theory for very good reasons.

\section{Problems with the minimal model}\label{problems}

While the minimal model was found to be very successful in fitting fermion data, it was realized soon thereafter that it faced some hurdles once the symmetry breaking constraints are included \cite{Aulakh:2005bd,Bajc:2005qe,Aulakh:2005mw,Bertolini:2006pe}.  The overall scale of the right-handed neutrino masses comes out to be $V_R \sim (10^{12} - 10^{13})$ GeV from fits to light neutrino masses, while consistency of symmetry breaking of the model requires $V_R \sim (10^{15}-10^{16})$ GeV. If $V_R$ takes values in the range of $(10^{12} - 10^{13})$ GeV, then certain colored multiplets from various Higgs fields will have masses of order $V_R^2/M_{\rm GUT} \sim 10^{10}$ GeV, spoiling perturbative gauge coupling unification.

There is an independent issue related to proton decay mediated by the color-triplet Higgsinos of the model.  The partial lifetime for the decay $p \rightarrow \overline{\nu} K^+$ would be shorter than the experimental lower limit of $5.9 \times 10^{33}$ years \cite{Hayato:1999az} if all the SUSY particles have masses of order TeV.  While this is not a problem in itself, as the SUSY particle masses are currently unknown, a realistic version of SUSY $SO(10)$ should also be compatible with this constraint.  Assuming TeV-scale superparticle masses, it has been proposed that this issue can be overcome by a cancelation mechanism with the inclusion of a $120_H$ \cite{Dutta:2004zh,Mohapatra:2018biy} into the minimal model.  Another possibility that has been suggested to suppress proton decay rate is via non-perturbative (but uncontrollable) suppressions \cite{Aulakh:2013lxa} with an asymptotically safe
theory in mind \cite{Bajc:2016efj}. Here we show explicitly the severity of this constraint in the minimal model (without a 
$120_H$), if all the SUSY particles have masses around a TeV.

Both of these problems, too small a scale for neutrino masses and too short a lifetime for the proton with TeV superparticles, could in principle be solved in a split-SUSY scenario \cite{Giudice:2004tc,ArkaniHamed:2004yi,Wells:2004di} with the gaugino--Higgsino masses
around $100$ TeV and the squark--slepton masses around $10^{13}$ GeV \cite{Bajc:2008dc}. This would allow for an increase in the light
neutrino masses, because the gauge couplings can be kept smaller for higher energies before the bosonic threshold is reached. Simultaneously,
$d=5$ proton decay rate would be highly suppressed, owing to the large sfermion
masses. Unfortunately, this solutions turns out to have various shortcomings, both theoretical and experimental.
First, the Higgs boson mass of $125$ GeV discovered after this proposal was made does not allow such a large SUSY breaking scale with small $A$-terms
\cite{Giudice:2011cg,Bagnaschi:2014rsa}. Second, the reactor neutrino mixing angle, although large, was not large enough in
\cite{Bajc:2008dc}.\footnote{It has to be said however, that in the fit of Ref. \cite{Bajc:2008dc} only the upper experimental limit has been
put on $\theta_{13}$, so a new $\chi^2$ fit with the measured value of $\theta_{13}$ may well give a better and acceptable value.}
Third, there may be also issues with stability of the spectrum with such large values of the bosonic masses \cite{Haba:2006kb}, although this may
be an issue of naturalness only.

\section{Proposed solution to the problems}

It is clear that the minimal SUSY $SO(10)$ model consisting of $\{10_H + \overline{126}_H + 126_H + 210_H\}$ Higgs fields needs to be extended, in view the problems it faces.  One avenue is to add a $120_H$ to the theory in which case the fermion mass fits of the model will be significantly affected.  Consistency has been shown in this case \cite{Dutta:2004zh,Mohapatra:2018biy,Aulakh:2008sn} (or in the non-supersymmetric case recently \cite{Babu:2016bmy}).  An alternative, which we discuss in this paper, is to modify the symmetry breaking sector without affecting the Yukawa sector.  This is achieved by adding a $54_H$, without the need for a $120_H$.  Thus the proposed model has the Higgs content of $\{10_H + \overline{126}_H + 126_H + 210_H+54_H\}$.

One can compare the number of parameters that are introduced by the addition of a $54_H$ with that for a $120_H$. Since $54_H$ has no Yukawa couplings to the fermions in $16$, the only new parameters are the 6 new complex couplings that appear in $W_{SSB}$.  After removing one phase by field redefinition of $54_H$, we arrive at  $26+12-1=37$ real parameters,  keeping the model still minimal among the renormalizable versions \cite{Aulakh:2003kg}.  In contrast, adding a $120_H$ (without a $54_H$) would introduce 3 complex Yukawa couplings and 9 new complex couplings in $W_{SSB}$, leading to $26+6+18-1 = 49$ real parameters.

The advantage of introducing a $54_H$ is that the symmetry breaking can proceed in three steps that is consistent with gauge coupling unification:
\begin{eqnarray}
SO(10) &\xrightarrow{\langle 54_H \rangle,\,\langle 210_H \rangle}& SU(3)_c \times SU(2)_L \times U(1)_Y  \times U(1)_{B-L} \nonumber \\
&\xrightarrow{\langle 126_H \rangle,\,\langle \overline{126}_H \rangle}& SU(3)_c \times SU(2)_L \times U(1)_Y \nonumber \\
&\xrightarrow{\langle 10_H \rangle}& SU(3)_c \times U(1)_{\rm em}
\end{eqnarray}
The first step of symmetry breaking occurs at $M_{GUT}\approx 10^{16}$ GeV, which leaves the rank of $SO(10)$ intact.  The next step occurs around $V_R \sim 10^{12}$ GeV where the  $U(1)_{B-L}$ symmetry gets spontaneously
broken. The $54_H$ allows to have such an intermediate scale without any light remnants, apart from a SM singlet pair.
The renormalization group running of the gauge couplings does not change except for a small threshold correction, which we found to be well
under control.    As mentioned in Sec. \ref{problems}, the requirement of  $V_R$ to be around $10^{12}$ GeV arises from fit to the fermion masses. The VEVs $\langle 126_H \rangle,\,\langle \overline{126}_H \rangle$  of this order is required to get the scale of the  light neutrino masses correct. Our numerical analysis performed in Sec. \ref{fit}, shows that the best fit value of $V_R$ corresponds to $\sim 5\times 10^{12}$ GeV. However, acceptable solutions to the fermion fit can be found which may deviate from this value by a factor of few. 

It is easy to see why with the addition of $54_H$ an intermediate scale $V_R$ for $B-L$ breaking is possible without any fields carrying SM quantum numbers surviving to $V_R$.  Note that the superpotential for the $210_H$ alone, viz., $W_{210_H} =\frac{1}{2}m_1210_H^2 +\lambda_1210_H^3$ allows for the breaking $SO(10) \rightarrow SU(5) \times U(1)$, while the superpotential for $54_H$ alone, viz., $W_{54_H} = m_554_H^2
+\lambda_854_H^3$ breaks $SO(10)$ down to $SU(4)_c \times SU(2)_L \times SU(2)_R$.  The joint effect of $210_H$ and $54_H$ is to break $SO(10)$ down to $SU(3)_c \times SU(2)_L \times U(1)_Y \times U(1)_{B-L}$.  The cross coupling in the superpotential, viz.,
$\lambda_{10}54_H210_H^2$, ensures that all of the would-be Goldstone bosons will acquire masses.  We have explicitly verified these statements by examining the full Higgs spectrum with the $54_H$ and with an intermediate scale $V_R \sim 10^{12}$ GeV. Such a solution is not permitted in the case of using $\{\overline{126}_H + 126_H + 210_H\}$ for GUT symmetry breaking due to the intricate nature of $\overline{126}_H-126_H-210_H$ cross coupling, which on the one hand should induce $SU(5)$-breaking VEV in $210_H$, and on the other hand should also supply masses for the would-be Goldstone bosons.

The problem of rapid proton decay via  $d=5$ operators can be kept under control only with a mini-split SUSY scenario. We will show
that a SUSY spectrum with TeV scale gauginos and Higgsinos and  $100$ TeV squark and sleptons would suppress the most dangerous mode sufficiently
to be in accord with the experimental limit. Such a mini-split SUSY spectrum is compatible with pure gravity mediated SUSY breaking \cite{Ibe:2011aa,Ibe:2012hu}, which we shall elaborate on.  Within such a framework, the partial lifetime for the decay $p \rightarrow \overline{\nu} K^+$ is found to be about an order of magnitude above the present experimental lower limit.

We perform a new fit to fermion data, taking into account the two new ingredients of this proposed scenario: a mini-split
SUSY case, and the extra intermediate $U(1)_{B-L}$ scale. We have found that an excellent fit to all light fermion masses and mixings is
possible in this scenario.

\section{The Higgs potential and the vacuum structure}

The supersymmetric Higgs sector of $SO(10)$ involving $\{10_H + \overline{126}_H + 126_H + 210_H+54_H\}$ has been studied in Ref. \cite{Fukuyama:2004ps}.  We shall follow closely the notation of that paper.
The most general renormalizable Higgs superpotential is:\footnote{The missing masses and couplings are
connected to $45_H$ and $120_H$ used in Ref. \cite{Fukuyama:2004ps}, which are, however, not part of our model.}
\bea
W_{SO(10)}^{SSB}&=&\frac{1}{2}m_1210_H^2+m_2\overline{126}_H126_H+m_310_H^2+\frac{1}{2} m_5 54_H^2\non\\
&+&\lambda_1210_H^3+\lambda_2210_H\overline{126}_H126_H+\lambda_3126_H10_H210_H
+\lambda_4\overline{126}_H10_H210_H\non\\
&+&\lambda_854_H^3+\lambda_{10}54_H210_H^2+\lambda_{11}54_H126_H^2+\lambda_{12}54_H\overline{126}_H^2
+\lambda_{13}54_H10_H^2
\label{fuku}
\eea

By denoting the vacuum expectation values in the Pati-Salam SU(4)$_c\times$SU(2)$_L\times$SU(2)$_R$
notation as
\bea
\langle(1,1,1)_{210}\rangle=V_1&,&
\langle(15,1,1)_{210}\rangle=V_2\;\;\;,\;\;\;
\langle(15,1,13)_{210}\rangle=V_3\\
\langle(1,1,1)_{54}\rangle=V_E&,&
\langle(\overline{10},1,3)_{126}\rangle=V_R\;\;\;,\;\;\;
\langle(10,1,3)_{\overline{126}}\rangle=\overline{V}_R
\eea
\noi
we get in the limit $|V_R|=|\overline{V}_R|<< V_{1,2,3,E}$ for the vacuum expectation values \cite{Babu:2016cri}
\bea
\label{V1}
\frac{V_1}{V_2}&=&\frac{1}{2\sqrt{3}}\left(\frac{V_3}{V_2}\right)^2 \label{SE-1}\\
\frac{m_1}{V_2}&=&-\frac{\lambda_1}{5\sqrt{2}}\left(3+\left(\frac{V_3}{V_2}\right)^2\right)\\
\frac{V_E}{V_2}&=&\frac{\lambda_1}{\lambda_{10}\sqrt{30}}\left(-2+\left(\frac{V_3}{V_2}\right)^2\right)\\
\label{m2}
\frac{m_2}{V_2}&=&-\frac{\lambda_2}{120}\left(6\sqrt{2}+12\left(\frac{V_3}{V_2}\right)+\sqrt{2}
\left(\frac{V_3}{V_2}\right)^2\right) \label{vr}\\
\frac{m_5}{V_2}&=&\frac{1}{20\sqrt{2}\lambda_1\lambda_{10}}\left(\left(-5\lambda_{10}^3-2\lambda_1^2\lambda_8\right)
\left(\frac{V_3}{V_2}\right)^2+\left(-20\lambda_{10}^3+4\lambda_1^2\lambda_8\right)\right) \label{SE-2}
\eea

\noi or, 
\bea
\label{V1b}
\frac{V_1}{V_2}&=&-\frac{\sqrt{3}}{2}\\
\frac{m_1}{V_2}&=&-\frac{\sqrt{2}\lambda_1}{30}\left(3+\left(\frac{V_3}{V_2}\right)^2\right)\\
\frac{V_E}{V_2}&=&\frac{\lambda_1}{\lambda_{10}\sqrt{30}}\left(1+2\left(\frac{V_3}{V_2}\right)^2\right)\\
\label{m2b}
\frac{m_2}{V_2}&=&-\frac{\lambda_2}{40}\left(\sqrt{2}+4\left(\frac{V_3}{V_2}\right)\right) \label{vrb}\\
\frac{m_5}{V_2}&=&-\frac{1}{20\sqrt{2}\lambda_1\lambda_{10}}\left(2\lambda_1^2\lambda_8
\left(\frac{V_3}{V_2}\right)^2+\left(5\lambda_{10}^3+2\lambda_1^2\lambda_8\right)\right) \label{SE-2b}
\eea

\noi
which spontaneously break $SO(10)$ down to $SM\times$U(1)$_{B-L}$, if $V_R = \overline{V}_R$ is ignored.  This symmetry breaks further down to the SM once a nonzero $V_R$ is generated.  Eq. (\ref{vr}) is the condition necessary for inducing a nonzero $V_R$.

The Higgs spectrum of this theory has been worked out in Ref. \cite{Fukuyama:2004ps}. It is easy to verify for this spectrum
that with $V_R \ll V_{1,2,3}, V_E$, there are no light states in the theory, except for two SM singlet fields $\sigma +
\overline{\sigma}$ carrying $B-L$ charges of $\pm2$.  These fields acquire VEVs at an intermediate scale, breaking the $B-L$
symmetry and thereby generating heavy right-handed neutrino masses.

The solutions of the stationary conditions given in Eqs. (\ref{V1})-(\ref{SE-2}) and Eqs. (\ref{V1b})-(\ref{SE-2b}) will be used in our numerical computations of proton decay amplitude.

\section{The Yukawa sector and GUT threshold corrections}

The renormalizable Yukawa superpotential of the model is given in Eq. (\ref{sup}).
After the GUT symmetry breaking parametrized by the VEVs in Eqs. (\ref{V1})-(\ref{m2}) or Eqs. (\ref{V1b})-(\ref{m2b}) the only light states (compared to $M_{GUT}$) are the SM
superfields and the SM singlets $\sigma$, $\bar\sigma$ which for small $V_R$ are purely from $126_H$,
$\overline{126}_H$ that do not mix with other singlets. As a result,
after integrating out all the heavy states with mass $M_{GUT}$, the superpotential below the GUT scale will have the form:
\beq
W=W_{Yukawa}+W_{Higgs}
\eeq
\noi
with
\begin{align}
W_{Yukawa}&=\frac{1}{2}Y_{R}^{ij}\nu^c_i\nu^c_j\overline{\sigma}+Y_{\nu_D}^{ij}\nu^c_iL_jH_u+Y_U^{ij}u^c_iQ_jH_u
+Y_D^{ij}d^c_iQ_jH_d+Y_E^{ij}e^c_iL_jH_d\\
\label{WHiggs}
W_{Higgs}&=\mu H_uH_d+\frac{V_R^2}{M_{GUT}}\sigma\bar\sigma-\frac{(\sigma\bar\sigma)^2}{2M_{GUT}}~.
\end{align}
Here the mass term for $\sigma$ field and the quartic coupling are determined in terms of the fundamental
parameters of Eq. (\ref{fuku}), but their actual forms are not so relevant for our analysis. After minimizing the potential we get $\left|\langle\sigma\rangle\right|=\left|\langle\bar\sigma\rangle\right|=V_R$. The Majorana Yukawa coupling matrix $Y_R$ obeys the
boundary condition $Y_R(M_{GUT}) = Y_{126}(M_{GUT})$.

At $M_{GUT}$ we have the following SO(10) relations:
\bea
v_uY_U&=&v_u^{10}Y_{10}+v_u^{126}Y_{126} \label{y1} \\
v_dY_D&=&v_d^{10}Y_{10}+v_d^{126}Y_{126}\\
v_dY_E&=&v_d^{10}Y_{10}-3v_d^{126}Y_{126}\\
v_uY_{\nu_D}&=&v_u^{10}Y_{10}-3v_u^{126}Y_{126}\\
M_N&=&v_LY_{126}-\left(v_uY_{\nu_D}\right)^T\left(V_RY_{126}\right)^{-1}\left(v_uY_{\nu_D}\right) \label{y2}~.
\eea
Here $v_u$ and $v_d$ are the VEVs of the MSSM fields $H_u$ and $H_d$, and we define as usual $\tan\beta = v_u/v_d$.  While the light neutrino masses receive contributions from type-I seesaw as well as type-II seesaw, the magnitude of the latter turns out to be small.  This is because the weak triplet(s) in the model have masses of order the GUT scale.  In our analysis we keep only the type-I contribution to neutrino masses.

Since $M_{\nu_R}$ and $V_R$ are less than $M_{GUT}$, we have corrections to
the 1-step RGE running from $M_Z$ to $M_{GUT}$. To analyze these effects, we choose without loss of generality a basis where $Y_{126}$ is real and diagonal.  The corrections to the Yukawa couplings due to the intermediate scale threshold is obtained by integrating the RGE equations, Eqs. \eqref{th1}-\eqref{th2} of Appendix \ref{D} (denoting the diagonal entries of $Y_{126}$ as $Y_{126}^d$):
\begin{align}
{{Y^\prime}_{\nu_D}}^{ij}={Y_{\nu_D}}^{ij}-\frac{1}{(4\pi)^2}\sum_{k=1}^3&\left[
3\left(Y_{\nu_D}Y_{\nu_D}^\dagger\right)^{ik}\left(Y_{\nu_D}\right)^{kj}
\log{\left(\frac{M_{GUT}}{max\left(\left|V_R\left(Y_{126}^d\right)^i\right|,\left|V_R\left(Y_{126}^d\right)^k\right|\right)}\right)}\right.\non\\
&\left.+{Y_{\nu_D}}^{ij}\left(Y_{\nu_D}Y_{\nu_D}^\dagger\right)^{kk}
\log{\left(\frac{M_{GUT}}{max\left(\left|V_R\left(Y_{126}^d\right)^i\right|,\left|V_R\left(Y_{126}^d\right)^k\right|\right)}\right)}\right]
\label {eq1} \non\\
-\frac{1}{(4\pi)^2}&\left[-\frac{3}{2} g_B^2{Y_{\nu_D}}^{ij}\log{\left(\frac{M_{GUT}}{g_B V_R}\right)}\right.\non\\
&\left.+\left|\left(Y_{126}^d\right)^i\right|^2\left(Y_{\nu_D}\right)^{ij}
\log{\left(\frac{M_{GUT}}{max\left(\left|V_R\left(Y_{126}^d\right)^i\right|,\left|V_R^2/M_{GUT}\right|\right)}\right)}\right]\\
{Y^\prime_E}^{ij}=Y_E^{ij}-\frac{1}{(4\pi)^2}&\left(\sum_{k=1}^3\left(Y_EY_{\nu_D}^\dagger\right)^{ik}\left(Y_{\nu_D}\right)^{kj}
\log{\left(\frac{M_{GUT}}{\left|V_R\left(Y_{126}^d\right)^k\right|}\right)}-  \frac{3}{2} g_B^2Y_E^{ij}\log{\left(\frac{M_{GUT}}{g_B V_R}\right)}\right)\\
{Y^\prime_U}^{ij}=Y_U^{ij}-\frac{1}{(4\pi)^2}&Y_U^{ij}\left(\sum_{k=1}^3\left(Y_{\nu_D}^\dagger Y_{\nu_D}\right)^{kk}
\log{\left(\frac{M_{GUT}}{\left|V_R\left(Y_{126}^d\right)^k\right|}\right)}-\frac{1}{6}g_B^2\log{\left(\frac{M_{GUT}}{g_BV_R}\right)}\right)\\
{Y^\prime_D}^{ij}=Y_D^{ij}-\frac{1}{(4\pi)^2}&Y_D^{ij}\left(-\frac{1}{6}g_B^2\log{\left(\frac{M_{GUT}}{g_BV_R}\right)}\right) \label{eq2}
\end{align}

\vspace{0.2in}
\noi
where $Y^\prime_{U,D,E}$  are the numerical values one would get by solving the RGEs from $M_Z$ to
$M_{GUT}$ without intermediate states.  On the other hand, the numerical value of $M_N$ with $Y_{\nu_D}$ replaced by $Y_{\nu_D}^{\prime}$   is obtained by solving the RGE from $M_Z$ to the intermediate scale.  We shall use these equations as the first step in obtaining fit to fermion masses and mixings.  This would determine, to a very good approximation, the intermediate spectrum as well as the scale $V_R$ from the fermion fit.  One can in principle  improve
the fermion fit by using this intermediate spectrum and using the exact one-loop RGE for crossing the threshold as given in Eqs. \eqref{th1}-\eqref{th2} of Appendix \ref{D}. However, our numerical analysis performed in the next section shows that these threshold effects are very small, hence we do not repeat the process.

\section{Fit to the fermion masses and mixings}\label{fit}

We rewrite Eqs. \eqref{y1} - \eqref{y2} as:
\begin{align}
&Y_D= H + F\\
&Y_U= r (H + s F)\\
&Y_E= H -3 F\\
&Y_{\nu_D}= r (H - 3 s F)\\
&M_N= -\left(v_uY_{\nu_D}\right)^T\left(c_R F\right)^{-1}\left(v_uY_{\nu_D}\right),
\end{align}
\noindent where we have defined
\begin{align}
& r= \frac{v^{10}_u}{v^{10}_d} \frac{1}{\tan\beta}, \;\;\; s= \frac{v^{126}_u}{v^{126}_d} \frac{v^{10}_d}{v^{10}_u}, \;\;\; c_R= V_R \frac{v_d}{v^{126}_d},  \label{rs-original} \\
&Y_{10}=\frac{v_d}{v^{10}_{d}} H, \;\;\;
 Y_{126}=\frac{v_d}{v^{126}_{d}} F.
\end{align}

\noindent As noted before, this Yukawa sector has 12 real parameters and 7 phases to fit 18 measured quantities.
Without loss of generality one can go to a basis where the symmetric matrix $F$ is real and diagonal, whence $H$ becomes a general complex symmetric matrix. The parameters $r$ and $c_R$ can be made real, whilst $s$ will remain a complex parameter.

 To fit the fermion masses and mixings we perform a $\chi^2$-analysis, with the pull and the $\chi^2$-function defined as:
\begin{align}
P_i=\frac{O_{i\;th}-E_{i\;exp}}{\sigma_i},\;
\chi^2= \sum_i  P^2_i,
\end{align}

\noindent
where $\sigma_i$ represent experimental $1\sigma$ uncertainty and $O_{i\;th}$ and  $E_{i\;exp}$ represent  theoretical prediction, experimental central value and pull of observable
$i$. We perform the fit  at the GUT scale, which we take to be
$M_{GUT}= 2\times 10^{16}$ GeV.

\FloatBarrier
\begin{table}[th!]
\centering
\footnotesize
\resizebox{0.9\textwidth}{!}{
\begin{tabular}{|c|c|c|c|}
\hline
Yukawa Couplings & Central Values  & \pbox{10cm}{CKM parameters \\ \& Neutrino Parameters} & Central Values  \\ [1ex] \hline
$y_{u}/10^{-6}$ & $6.65 \pm 2.25$ &$\theta^{\rm{CKM}}_{12}$ & $0.22735\pm 0.00072$  \\ \hline
$y_{c}/10^{-3}$ & $3.60 \pm 0.11$ &$\theta^{\rm{CKM}}_{23}/10^{-2}$ & $4.208\pm 0.064$    \\ \hline
$y_{t}$ & $0.9860 \pm 0.00865$ &$\theta^{\rm{CKM}}_{13}/10^{-3}$ & $3.64\pm 0.13$   \\ \hline
$y_{d}/10^{-5}$ & $1.645 \pm 0.165$ &$\delta^{\rm{CKM}}$ & $1.208\pm0.054$  \\ \hline
$y_{s}/10^{-4}$ & $3.125 \pm 0.165$ &$\Delta m^{2}_{21}/10^{-5} eV^{2}$ &7.56$\pm$0.19  \\ \hline
$y_{b}/10^{-2}$ & $1.639 \pm 0.015$ &$\Delta m^{2}_{31}/10^{-3} eV^{2}$ &2.55$\pm$0.04  \\ \hline
$y_{e}/10^{-6}$ & $2.79475 \pm 0.0000155$ &$\sin^{2}\theta^{\rm{PMNS}}_{12}/10^{-1}$  &3.219$\pm$0.17  \\ \hline
$y_{\mu}/10^{-4}$ & $5.89986 \pm 0.0000185$ &$\sin^{2}\theta^{\rm{PMNS}}_{23}/10^{-1}$ &4.31$\pm$0.19 \\ \hline
$y_{\tau}/10^{-2}$ & $1.00295 \pm 0.0000905$  &$\sin^{2}\theta^{\rm{PMNS}}_{13}/10^{-2}$ &2.1625$\pm$0.0825  \\ \hline
\end{tabular}
}
\caption{  Values of  observables in the charged fermion sector at $M_{Z}$ scale, taken from Ref. \cite{Antusch:2013jca}. Here  experimental central
values with associated 1 $\sigma$ uncertainties are quoted. The masses of the charged  fermions are given by the relations $m_i = v \;y_i$
with $v=174.104$ GeV. For neutrino observables, the low energy values are taken from Ref. \cite{deSalas:2017kay}.}
\label{input}
\end{table}

To get the GUT scale values
of the observables in the charged fermion sector we take the central values at the $M_Z$
scale from Table-1 of Ref. \cite{Antusch:2013jca}. For neutrino observables, the low energy values are taken from Ref. \cite{deSalas:2017kay}. For reader's convenience, we collect all these low energy values in Table  \ref{input}.  With these inputs, we do the RGE running of the Yukawa couplings \cite{Machacek:1983fi,Arason:1991ic}, the CKM parameters \cite{Babu:1987im} and the effective couplings of the neutrino 5D operator, $\kappa$ \cite{Babu:1993qv,Chankowski:1993tx,Antusch:2001ck} within the SM up to the scale $\mu=$1 TeV.  Proton decay constraints on the model requires a mini-split SUSY scenario, wherein
the gauginos and Higgsinos are taken to be at the TeV scale in order to satisfy gauge coupling unification constraint as well the dark matter constraints. Proton decay constraints lead to a lower limit on the squark and slepton masses of  $\mathcal{O}(100)$ TeV within the model. For the fit purpose, we fix the squark and slepton mass scale to be 100 TeV. In order to correctly take into account the  evolution of the observables, we run the modified RGEs in between $\mu=$1 TeV and $\mu=$100 TeV using split-SUSY RGEs. We collect the relevant RGEs in Appendix \ref{C}. Above 100 TeV, the full MSSM is restored, so we use the  corresponding MSSM RGEs \cite{Barger:1992ac, Barger:1992pk, Antusch:2001vn} and evolve them up to the GUT scale, $M_{GUT}$. For the charged fermion observables, we fit to these evolved values at the GUT scale. However, note that the first stage symmetry breaking step is $SO(10)\to SM \times U(1)_{B-L}$, hence, threshold corrections due to the right handed neutrinos and the $U(1)_{B-L}$ gauge boson need to be taken into account from the $V_R$ scale to the GUT scale. So above the SUSY scale,  the effective couplings of the neutrino 5D operator running is performed up to the intermediate scale instead of the GUT scale.  The relevant RGEs for the Yukawa couplings in between these energy scales ($V_R$ and $M_{GUT}$) are given in Appendix \ref{D} and the corresponding corrected Yukawa couplings derived from these equations  are given in  Eqs. \eqref{eq1}-\eqref{eq2}.  We do the fitting of the observables at the GUT scale using these threshold corrected Yukawa couplings.

The values of the observables resulting from the RGE running (the input values) and also the best fit values  are presented in  Table \ref{result}.     The best fit parameters corresponding to $\tan\beta=10$ are found to be:
\begin{align}\label{rs-values}
 r=8.730612,\;s= 3.397119\times 10^{-1}+1.039406\times 10^{-2}\;i, \;c_R= 3.187055\times 10^{14} \;\rm{GeV}.
\end{align}
\begin{align}\label{H-value}
H=
\left(
\scalemath{0.7}{
\begin{array}{ccc}
8.242157\times 10^{-5} - 9.797097\times 10^{-5} \; i&
-5.245430\times 10^{-4}  + 4.580976\times 10^{-4}  \; i&
2.334517\times 10^{-3}  + 1.273798\times 10^{-3}  \; i\\
-5.245430\times 10^{-4}  + 4.580976\times 10^{-4}  \; i&
2.117891\times 10^{-3} - 2.129217\times 10^{-3}  \; i&
-1.111526\times 10^{-2}  - 7.542882\times 10^{-3}  \; i\\
2.334517\times 10^{-3}  + 1.273798\times 10^{-3}  \; i&
-1.111526\times 10^{-2}  - 7.542882\times 10^{-3}  \; i&
-1.649528\times 10^{-2} + 5.158391\times 10^{-2} \; i
\end{array}
}
\right),
\end{align}
\begin{align}\label{F-value}
F=
\left(
\scalemath{0.8}{
\begin{array}{ccc}
4.052595\times 10^{-5}&0&0 \\
0&1.963342\times 10^{-3}&0\\
0&0&1.316235\times 10^{-2}
\end{array}
}
\right).
\end{align}

We shall use these best fit parameters in our evaluation of proton lifetime, discussed in the next section.

\FloatBarrier
\begin{table}[!b]
\centering
\footnotesize
\resizebox{0.7\textwidth}{!}{
\begin{tabular}{|c|c|c|c|}
\hline
\pbox{10cm}{Masses (in GeV) and \\  Mixing parameters} & \pbox{10cm}{~~~~~Inputs \\ (at $\mu= M_{GUT}$)} & \pbox{25cm}{Fitted values \\ (at $\mu= M_{GUT}$)} & \pbox{10cm}{~~~pulls}   \\ [1ex] \hline
$m_{u}/10^{-3}$ & 0.450$\pm$0.139 & 0.454 & 0.028 \\ \hline
$m_{c}$   & 0.248$\pm$0.007 & 0.245 & 0.175  \\ \hline
$m_{t}$   & 84.53$\pm$0.84 & 84.49 & -0.057 \\ \hline
$m_{d}/10^{-3}$  & 0.951$\pm$0.19 & 0.585 & -1.92  \\ \hline
$m_{s}/10^{-3}$  & 18.07$\pm$0.97 & 18.46 & 0.409   \\ \hline
$m_{b}$  & 0.961$\pm$0.009 & 0.961 & 0.048  \\ \hline
$m_{e}/10^{-3}$   & 0.379457 & 0.379468 & 0.002  \\ \hline
$m_{\mu}/10^{-3}$  & 80.1068 & 80.0416 & -0.081  \\ \hline
$m_{\tau}$  & 1.36781 & 1.36799 & 0.012   \\ \hline
$|V_{us}|/10^{-2}$  & 22.54$\pm$0.06 & 22.54 & 0.057  \\ \hline
$|V_{cb}|/10^{-2}$  & 4.14$\pm$0.06 & 4.14 & 0.013  \\ \hline
$|V_{ub}|/10^{-2}$  & 0.358$\pm$0.012 & 0.358 & 0.020 \\ \hline
$\delta_{CKM}$ & 1.208$\pm$0.054  & 1.222 & 0.265  \\ \hline
$\Delta m^{2}_{sol}/10^{-5}$(eV$^{2}$) & 8.679$\pm$0.218 & 8.683 & 0.019 \\ \hline
$\Delta m^{2}_{atm}/10^{-3}$(eV$^{2}$) & 2.929$\pm$0.046 & 2.929 &- 0.011 \\ \hline
$\sin^{2}\theta^{\rm{PMNS}}_{12}$ & 0.3219$\pm$0.017 & 0.3204 & -0.029  \\ \hline
$\sin^{2}\theta^{\rm{PMNS}}_{23}$ & 0.431$\pm$0.019 & 0.4281 & -0.0148  \\ \hline
$\sin^{2}\theta^{\rm{PMNS}}_{13}$ & 0.0216$\pm$0.00082 & 0.02148 & -0.145 \\ \hline
\end{tabular}
}
\caption{ Best fit values of the observables corresponding to the  type-I dominance seesaw scenarios is presented. This best fit corresponds to total $\chi^2=4$. The values of the observables at the GUT scale are obtained by RGE evolution as explained detail in the text. We have chosen $M_{GUT}= 2\times 10^{16}$ GeV, $g_B= 0.708$ and we have fixed $\tan\beta= 10$\; for this numerical analysis.  For the associated $1\;\sigma$ uncertainties of the observables at the GUT scale, we keep the same percentage uncertainty with respect to the central value of each quantity as that at the $M_Z$ scale. For the charged lepton Yukawa couplings at the GUT scale, a relative uncertainty of $1\%$ is assumed in order to take into account the theoretical uncertainties arising for example from threshold effects. As explained in the text, the RGE running of the two mass squared differences in the neutrino sector are performed from the low scale to the intermediate scale.  The parameters corresponding to this best fit are presented in Eqs. \eqref{rs-values} - \eqref{F-value}.     }
\label{result}
\end{table}

\vspace{10pt}
\noindent
\textbf{Prediction of the model for neutrino CP violation}
\vspace{5pt}

\begin{figure}[tbh]
\begin{center}
\includegraphics[width=12.5cm]{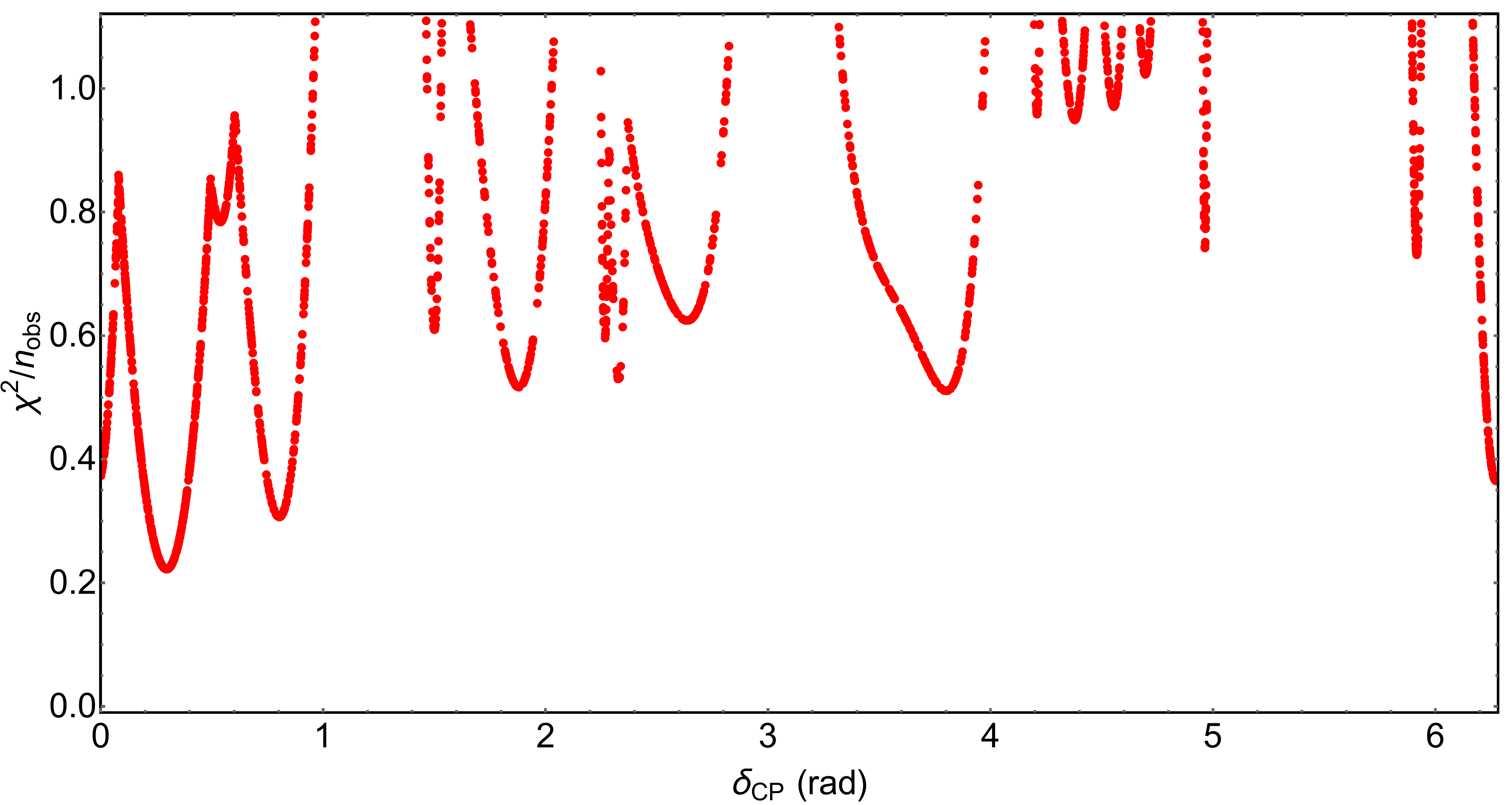}
\end{center}
\caption{\label{del} Variation of the Dirac type CP violating phase $\delta_{CP}$ in the neutrino sector by marginalizing over all other model parameters. For this plot we restrict ourselves to the case of minimum of the total $\chi^2 \leq 20$ (for 18 observables).  }
\end{figure}

Corresponding to the best fit, the predicted quantities in our model are presented in Table \ref{pred}.  We note that the best fit value of the CP violating parameter $\delta_{CP}$ for neutrino oscillations is 17$^0$, which is not however a firm prediction.
 In Fig. \ref{del}, we show the variation of this phase  by marginalizing over all other model parameters. This plot   corresponds to the case of minimum value of the total $\chi^2 \leq 20$ (for 18 observables), which should all be acceptable.  The figure shows the presence of multiple local minima with the global minimum corresponds to $\delta_{CP}=17^{\circ}$ with $\chi^2$ per degree of freedom equal to 0.2.  For example, if $\delta_{CP}$ is measured to be $3\pi/2$, the model is still acceptable,  with a local minimum that corresponds to  $\chi^2$ per degree of freedom around 0.9.  

\FloatBarrier
\begin{table}[th]
\centering
\footnotesize
\resizebox{0.7\textwidth}{!}{
\begin{tabular}{|c|c|}
\hline
Quantity & \pbox{10cm}{Predicted Value }  \\ [1ex] \hline
$\{m_{1}, m_{2}, m_{3} \}$ (in eV) & $\{ 3.32\times 10^{-3}, 9.89\times 10^{-3}, 5.42\times 10^{-2} \}$   \\ \hline

$\{\delta^{PMNS}, \alpha^{PMNS}_{21}, \alpha^{PMNS}_{31} \}$ & $\{ 17.0^{\circ}, 344.13^{\circ}, 337.45^{\circ} \}$  \\ \hline

$\{m_{cos}, m_{\beta}, m_{\beta \beta} \}$ (in eV) & $\{ 6.74\times 10^{-2}, 6.47\times 10^{-3}, 6.11\times 10^{-3} \}$  \\ \hline

$\{M_{1}, M_{2}, M_{3} \}$ (in GeV)  & $\{ 1.29\times 10^{10}, 6.25\times 10^{11}, 4.13\times 10^{12} \}$ \\ \hline
\end{tabular}
}
\caption{ Predictions corresponding to the best fit values presented in Table \ref{result} for type-I dominance seesaw scenario.  $m_{i}$ are the light neutrino masses, $M_{i}$ are the right handed neutrino masses, $\alpha_{21,31}$ are the Majorana phases following the PDG parametrization, $m_{cos}=\sum_{i} m_{i}$, $m_{\beta}=\sum_{i} |U_{e i}|^{2} m_{i}$ is the effective mass parameter for beta-decay and $m_{\beta \beta}= | \sum_{i} U_{e i}^{2} m_{i} |$ is the effective mass parameter for neutrinoless double beta decay.}
\label{pred}
\end{table}

\section{Proton decay calculation}

At this point we can estimate the proton decay rate or, better, we can determine the minimal allowed value of the 
sfermion mass (assumed here
for simplicity to be universal) from the proton decay constraint. We assume that these rates are dominated by wino exchange
and take as a benchmark the value of its mass to be\footnote{One can easily transform the result for other
values of this mass, knowing that $\tau_p\propto1/m_{wino}^2$. } $m_{wino}=1$ TeV.  Different fits are possible and the resulting sfemions mass scale $m_S$ depends very much on that.

To compute proton decay  rate, we define the relevant amplitude functions $A^{\nu}_{ijk\rho}$ as \cite{Babu:1998wi}: 
\begin{align}
A^{\nu}_{1bc\rho}= \frac{M^{-1}_{T\;11}}{N^2_d} \left( \hat{A}^{\nu}_{1bc\rho}[H^{(1)},H^{(2)}] + x  \hat{A}^{\nu}_{1bc\rho}[F^{(1)},F^{(2)}] + y \hat{A}^{\nu}_{1bc\rho}[H^{(1)},F^{(2)}] + z  \hat{A}^{\nu}_{1bc\rho}[F^{(1)},H^{(2)}] \right), \label{xyz}
\end{align}

\noindent
where (denoting $Y^{(i)}=H^{(i)}$ or $F^{(i)}$)  
\begin{align}\label{AMP}
\hat{A}^{\nu}_{1bc\rho}[Y^{(1)},Y^{(2)}]=&
\sum^3_{\ell=1} Y^{(1)}_{11}Y^{(2)}_{2\ell}V_{1b}V_{2c}U_{\ell \rho}
- \sum^3_{\ell=1} Y^{(1)}_{22}Y^{(2)}_{1\ell}V_{2b}V_{2c}U_{\ell \rho}
-\sum^3_{\ell=1} Y^{(1)}_{12}Y^{(2)}_{1\ell}V_{1b}V_{2c}U_{\ell \rho}
\nonumber \\ &+
\sum^3_{\ell=1} Y^{(1)}_{12}Y^{(2)}_{2\ell}V_{2b}V_{2c}U_{\ell \rho}
-\sum^3_{\ell=1} Y^{(1)}_{13}Y^{(2)}_{1\ell}V_{1b}V_{3c}U_{\ell \rho}
+ \sum^3_{\ell=1} Y^{(1)}_{13}Y^{(2)}_{2\ell}V_{3b}V_{2c}U_{\ell \rho}
\nonumber \\ &+
\sum^3_{\ell=1} Y^{(1)}_{13}Y^{(2)}_{3\ell}V_{3b}V_{3c}U_{\ell \rho} - \sum^3_{\ell=1} Y^{(1)}_{23}Y^{(2)}_{1\ell}V_{3b}V_{2c}U_{\ell \rho}
-\sum^3_{\ell=1} Y^{(1)}_{23}Y^{(2)}_{1\ell}V_{2b}V_{3c}U_{\ell \rho}
\nonumber \\ & -
\sum^3_{\ell=1} Y^{(1)}_{33}Y^{(2)}_{1\ell}V_{3b}V_{3c}U_{\ell \rho}
+ \sum^3_{\ell=1} Y^{(1)}_{11}Y^{(2)}_{3\ell}V_{1b}V_{3c}U_{\ell \rho}
+ \sum^3_{\ell=1} Y^{(1)}_{12}Y^{(2)}_{3\ell}V_{2b}V_{3c}U_{\ell \rho},
\end{align}

\noindent
and the $x,y,z$ parameters  are defined as \cite{Goh:2003nv}:
\begin{align}
x=\left(\frac{\sqrt{3}}{q^{\ast}_2}\right)^2 \frac{M^{-1}_{T\;32}}{M^{-1}_{T\;11}}, \;\;\;
y=\left(\frac{\sqrt{3}}{q^{\ast}_2}\right) \frac{M^{-1}_{T\;12}}{M^{-1}_{T\;11}},  \;\;\;
z=\left(\frac{\sqrt{3}}{q^{\ast}_2}\right) \frac{M^{-1}_{T\;31}}{M^{-1}_{T\;11}}. \label{xyzdef}
\end{align}

\begin{figure}[th]
\begin{center}
\includegraphics[width=8.cm]{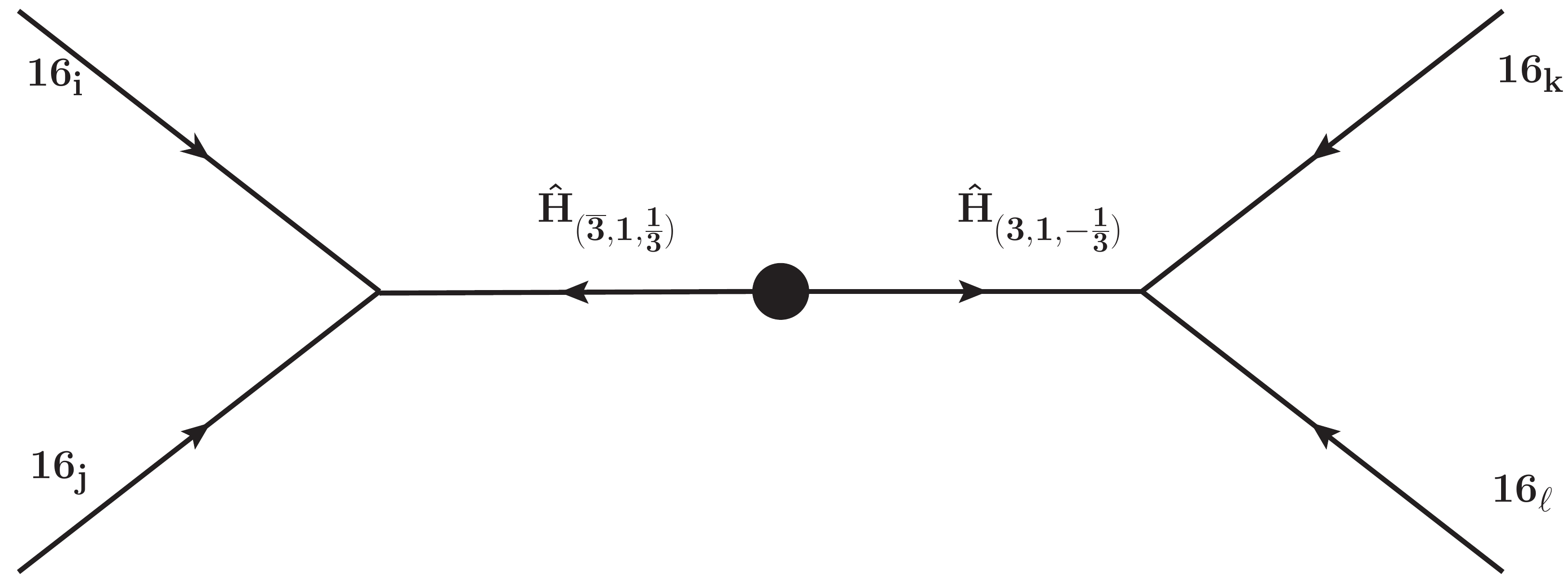}~~~~~~
\includegraphics[width=8.cm]{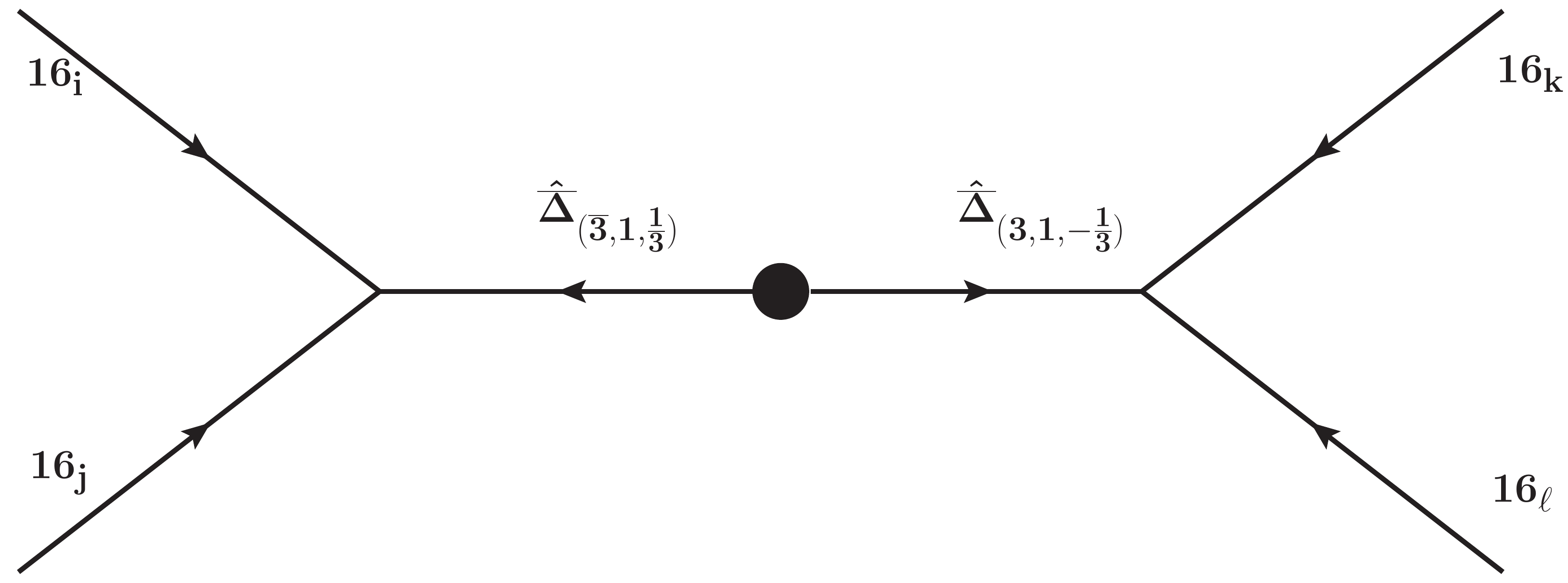}\\ \vspace{0.25in}
\includegraphics[width=8.cm]{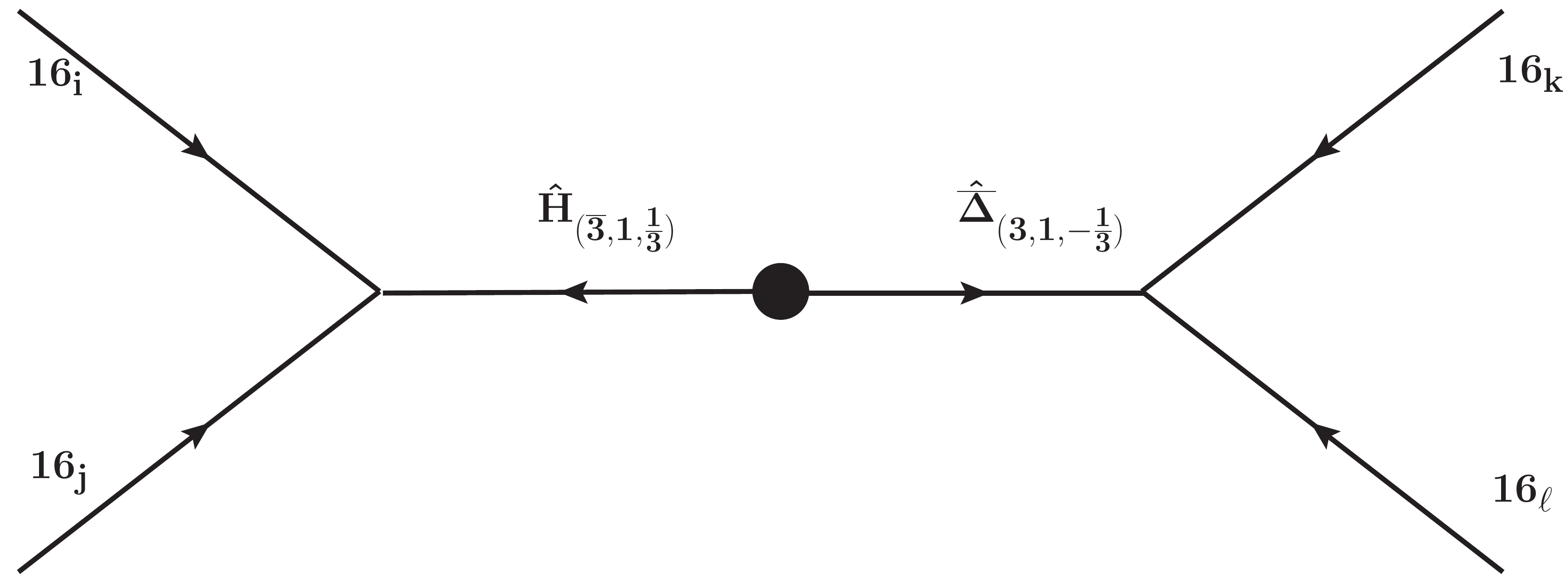}~~~~~~~
\includegraphics[width=8.cm]{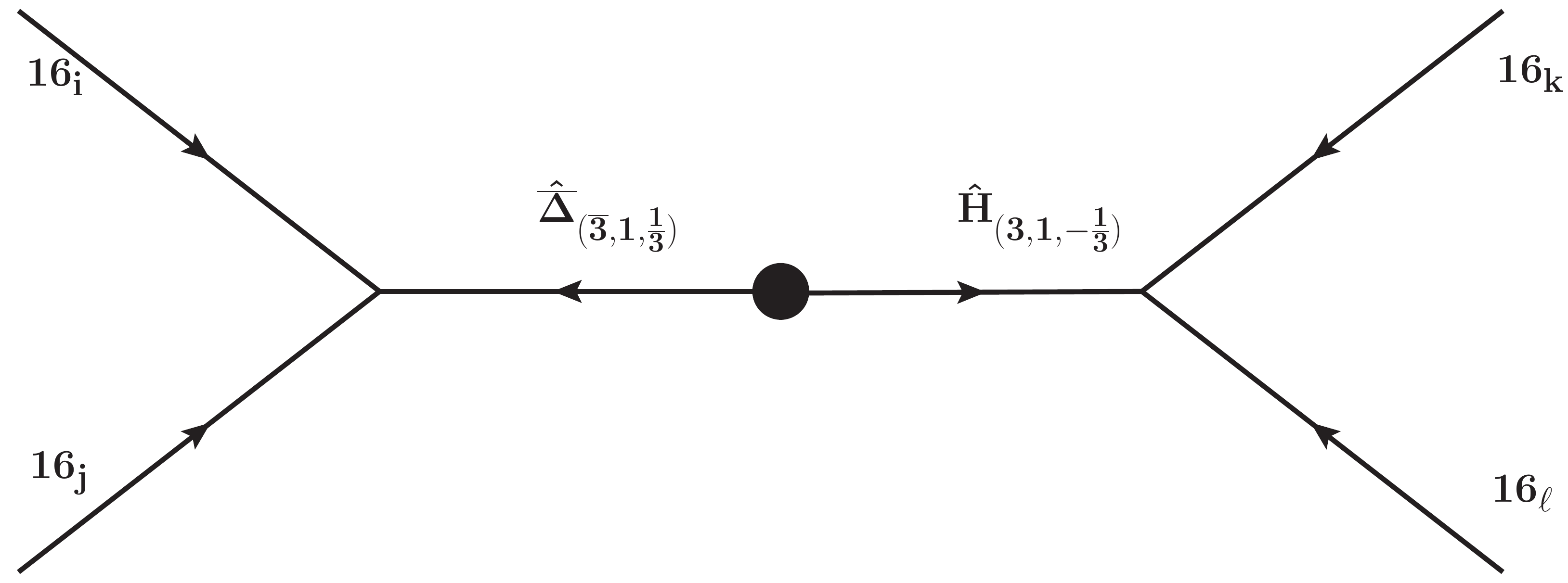}
\end{center}
\caption{\label{histogram} Supergraphs leading to $d=5$ proton decay operators.}
\label{feyn}
\end{figure}

\noindent Here, $M_T$ is the mass matrix for the color triplets which mediate the $d=5$ proton decay given in Eq. (\ref{triplet}), and $q_2$ is the component of MSSM field $H_d$ arising from $\overline{126}_H$ given in Eq. \eqref{q2} of Appendix \ref{B} (the normalization factor $N_d$ is also defined in Appendix \ref{B}). The supergraphs generating these amplitudes are shown in Fig. \ref{feyn}.

\FloatBarrier

The $H$ and $F$ matrices corresponding to the best fit are given in Eqs. \eqref{H-value}-\eqref{F-value} are clearly in a basis where the  $Y_{126}$  is diagonal. While calculating the proton decay amplitude functions, one needs to work in the physical basis of the particles. Denoting $u^{(g)}=V_u u^{(m)}$, where $(g)$ and $(m)$ stand for the gauge and the mass eigenstates (the corresponding matrices  for the down-quarks is $V_d$, charged leptons is $V_{\ell}$ and neutrinos is $V_{\nu}$), we have, $H^{(1)}= V^T_u H V_u, 
H^{(2)}= H^{(1)} V^{\prime}$ and $F^{(1)}= V^T_u F V_u, 
F^{(2)}=  F^{(1)} V^{\prime}$ (where $V^{\prime}= V_u^{\dagger}V_{\ell}$) \footnote{ For generality, in Eqs. \eqref{xyz} and \eqref{AMP} we kept $H^{(1,2)}$ and $F^{(1,2)}$, however, the result is in
our case independent on $V^\prime$, that is,  even if we use only $H^{(1)}$ and
$F^{(1)}$ all the time, the result would not change.}. Then, $V=V_{CKM}$ and  $U=U_{PMNS}$, here $V_{CKM}$ and $U_{PMNS}$ are the general mixing matrices that contain all the relevant phases. It is straightforward to compute all these matrices from the best fit values given in Eqs. \eqref{H-value}-\eqref{F-value}.

\newpage
The rates for the dominant proton decay processes in our model are given by: 
\begin{align}
\Gamma(p\rightarrow \overline{\nu}K^+) = & \frac{(m^2_p-m^2_K)^2 R^2_L}{32 \pi m^3_p}
  \left(\frac{g^2_2}{16 \pi^2} 2 f(m_{S},m_{S}) A_S\right)^2
\nonumber \\ & \times
 \sum_{\rho=1}^3 |A^{\nu}_{121\rho} \langle K^+|(us)_Ld_L|p\rangle  + A^{\nu}_{112\rho} \langle K^+|(ud)_Ls_L|p\rangle  |^2 ,
 \label{p1}
\end{align}

\begin{align}
\Gamma(p\rightarrow \overline{\nu}\pi^+) &= \frac{(m^2_p-m^2_{\pi})^2 R^2_L}{32 \pi m^3_p}
 \left(\frac{g^2_2}{16 \pi^2} 2 f(m_{S},m_{S}) A_S\right)^2
 \times
 \sum_{\rho=1}^3 |A^{\nu}_{111\rho} \langle \pi^+|(ud)_Ld_L|p\rangle    |^2 . \label{p2}
\end{align}

\noindent
The nuclear matrix elements appearing in Eqs. (\ref{p1})-(\ref{p2}) are given in Appendix \ref{B}, Eq. (\ref{nuc}). In general, the values of the dressing functions $f(a,b)$ are given by:
\beq
f(a,b)=
\left\{
\begin{array}{lcr}
1/2 & , &  a=b=1\\
\frac{1-b+b\log{b}}{(b-1)^2}& , & a=1\ne b\\
\frac{1-a+a\log{a}}{(a-1)^2}& , & b=1\ne a\\
\frac{-1+a-\log{a}}{(a-1)^2}& , & a=b\ne 1\\
\frac{1}{a-b}\left(\frac{a}{a-1}\log{a}-\frac{b}{b-1}\log{b}\right)& , &1\ne a\ne b\ne 1
\end{array}
\right.
\eeq

\noi
We define
\bea
\left(f_{UE}\right)_{ij}&=&f\left(\left(m^U_i\right)^2,\left(m^E_j\right)^2\right)\\
\left(f_{UD}\right)_{ij}&=&f\left(\left(m^U_i\right)^2,\left(m^D_j\right)^2\right),\\
\eea
which we take for simplicity to be universal ($m_S=m_{susy}$):
\beq
m^{U,D,E}_{1,2,3}=\frac{m_{susy}}{m_{wino}}.
\eeq

It is tempting to maximize the proton lifetime with respect to the parameters $(x,y,z)$ appearing in Eq. (\ref{xyz}), as was done in Ref. \cite{Goh:2003nv}.  However, we found that maximizing proton lifetime with respect to $(x,y,z)$ does not maximize it with the full set of
model parameters.  In particular, $M^{-1}_{T 11}/N^2_d$ appears in the amplitude, which should be included in the maximization process. The constraints
we obtain are much stronger than the ones quoted in Ref. \cite{Goh:2003nv}.

To get the constraints on the SUSY scale from proton decay  we followed the following procedure:

\bet

\item
We first set $\lambda_1=1$ and  $V_3=M_{GUT}=2\times10^{16}$ GeV. Then we randomly chose the perturbative values 
of the real input parameters $\lambda_2$, $\lambda_3$, $\lambda_{10}$, and of the complex input parameters $\lambda_4$, $\lambda_{11}$, 
$\lambda_{12}$, $\lambda_{13}$. Finally we chose complex $V_2$ of the same order as $V_3$. The equations of motion 
determine then $V_1$, $V_E$, $m_1$ and $m_2$ through Eqs. (\ref{V1})-(\ref{m2})  or Eqs. (\ref{V1b})-(\ref{m2b}), while the doublet-triplet fine-tuning 
determines $m_3$ and Eq. (\ref{SE-2}) or Eq. (\ref{SE-2b}) the parameter $m_5$ as a function of $\lambda_5$, which does not enter in any 
other quantity and is thus free. We demanded that the choice of parameters reproduce $r$ and $s$
(with at most $10\%$ error) given in  Eq. \eqref{rs-values};

\item
We took the Yukawa couplings from Eqs. \eqref{H-value}-\eqref{F-value}  and calculated the proton decay amplitude.

\item
With parameters so chosen, we calculated the minimal $m_S$ (sfermion mass) which satisfies all bounds for
proton decay.

\eet 
By taking  $\tau_p(p\to K^+\bar\nu)\geq5.9\times10^{33}$ yrs. and $\tau_p(p\rightarrow \overline{\nu}\pi^+)\geq3.9\times10^{32}$ yrs \cite{Hayato:1999az},
we got the histogram probability vs minimal $m_S$ (in GeV) shown in Fig. \ref{histogram} for 14478 fits. The minimal value among all fits obtained is $m_S=242.152$ TeV, for the inputs $|V_R|=|\overline{V}_R|=10^{12}$ GeV and
\bea
m_1/M_{GUT}&=&0.200704 +0.271952i\\
m_2/M_{GUT}&=& 0.104967 -0.144038i\\
m_3/M_{GUT}&=& 7.66344 +1.62961i
\eea
\bea
\lambda_1=1.&,&\lambda_{10}=0.656605\\
\lambda_2=-1.99&,&\lambda_{11}=-0.325041+0.986721i\\
\lambda_3=-1.47644&,&\lambda_{12}=-1.90723-0.44593i\\
\lambda_4=0.913794 +1.77904i&,&\lambda_{13}=0.877724 +0.840974i
\eea

\noi
while $\lambda_8$ can be chosen arbitrarily with $m_5$ given by eq. (\ref{SE-2}).

For these choices we get
\bea
V_1/M_{GUT}&=&1.16391 +1.7641i\\
V_2/M_{GUT}&=&-1.34397-2.037i\\
V_3/M_{GUT}&=&1.\\
v_E/M_{GUT}&=&-0.499196-0.376197
\eea
and
\beq
r=7.85755\;\;\;,\;\;\;s=0.306071 +0.00555256i
\eeq
which is within $10\%$ of the original $r,s$ from (\ref{rs-values}).

We thus conclude that with a mini-split SUSY spectrum, the model is compatible with proton lifetime constraints.

\begin{figure}[htb]
\begin{center}
\includegraphics[width=10.cm]{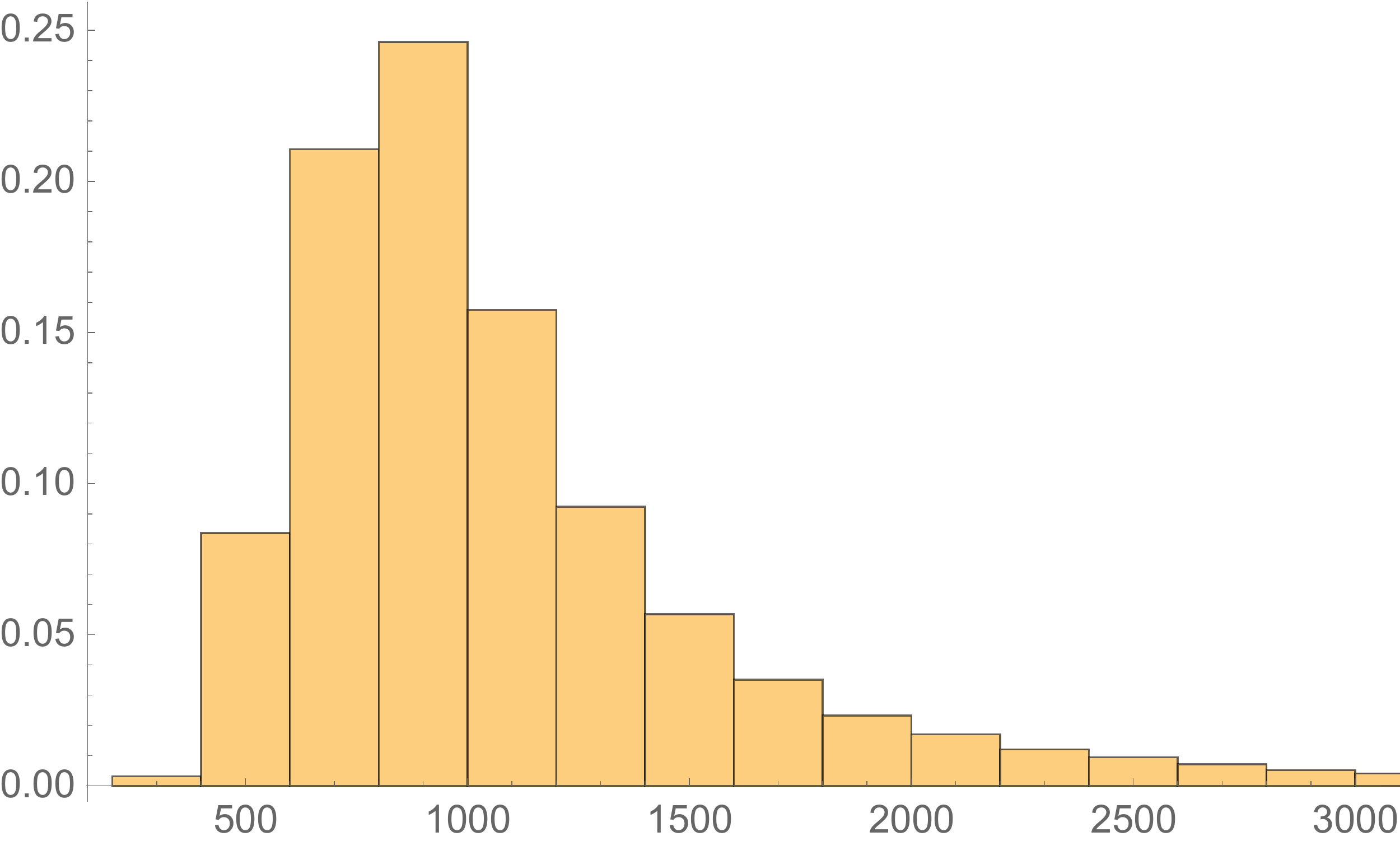}
\end{center}
\caption{\label{histogram} Histogram (the integral is normalized to 1) of the sfermion mass $m_S$ in TeV for 14478 different fits consistent with proton lifetime limits.}
\end{figure}

\section{Embedding of the model in pure gravity mediation}

As shown by the proton decay analysis, the model requires a heavy SUSY particle spectrum, with lighter gauginos and Higgsinos.  The latter will provide a dark matter candidate if the lightest of these fermions mass is of order TeV. While the squark and slepton masses can be varied at will without affecting gauge coupling unification (as they belong to complete multiplets of GUT), the same is not true for the gauginos and to some extend the Higgsinos.  Their masses being at the TeV scale would preserve gauge coupling unification as in the MSSM.

Such a mini-split SUSY spectrum may have a natural origin in pure gravity mediated SUSY breaking \cite{Ibe:2011aa,Ibe:2012hu}.  In this scheme, all mass parameters, including the $\mu$ term for the MSSM Higgsino mass, arise from SUSY breaking mediated by gravity.  Gaugino masses are zero at tree-level, but are induced by anomaly mediated contributions, as well as by Higgsino threshold effects.  Thus, these models have naturally light gauginos.  The mass spectrum of the gauginos is given by \cite{Ibe:2012hu}:
\begin{eqnarray}
M_1 &=& \frac{33}{5} \frac{g_1^2}{16 \pi^2} (m_{3/2} + \frac{1}{11}L) \nonumber \\
M_2 &=& \frac{g_2^2}{16\pi^2}(m_{3/2}+L) \nonumber \\
M_3 &=& -3 \frac{g_3^2}{16\pi^2} m_{3/2},
\end{eqnarray}
where
\begin{equation}
L = \mu_H \sin2\beta \frac{m_A^2}{|\mu_H|^2-m_A^2} {\rm ln} \frac{|\mu_H|^2}{m_A^2}~.
\end{equation}
Here terms proportional to $L$ arise from Higgsino threshold. The $\mu$ term and the $B\mu$ term in this scheme have the form
\begin{eqnarray}
\mu_H &=& c m_{3/2} \nonumber \\
B \mu_H &=& c m_{3/2}^2 + c' \frac{|F_X|^2}{ M_{Pl}^2}
\end{eqnarray}
where $c, c'$ are order one coefficients.  Thus the Higgsino threshold corrections to the gaugino masses are of the same order
as the anomaly mediated corrections.  Consequently, the mass ratios $M_2/M_3$ is not exactly predicted.

It should be noted that the neutral Wino is the lightest SUSY particle in this scheme, and can serve as the dark matter.  For a thermal dark matter, relic abundance would require the Wino mass $M_2$ to be near 2.7 TeV.  However, in these models, Wino dark matter can be produced non-thermally, via the decay of the gravitino, in which case the 2.7 TeV mass constraint does not hold.  In fact, the preferred range for Wino dark matter is below 1 TeV \cite{Ibe:2011aa}.  The Wino cannot be below about 200 GeV, as that would modify BBN significantly.

With Wino mass in the range 200 GeV to 1 TeV, the gluino can have mass of order $(3-20)$ TeV.  The squark and slepton masses are of order $(100 - 300)$ TeV. This is precisely the spectrum preferred in the model from proton decay constraints. Within this scenario, the squark and slepton masses cannot be much above 300 TeV, as opposed to general split-SUSY models.  This leads to a prediction for the partial lifetime for $p \rightarrow \overline{\nu} K^+$ within the model:  It cannot exceed a factor of 10 compared to the current lower limit of $5.9 \times 10^{33}$ yrs.  Thus the model is testable in proton decay searches.  However, if one deviates from this pure gravity mediation scenario, where the sfermion masses can be made very high, which would be consistent with gauge coupling unification,  such as in general split-SUSY scenario, proton decay rate will be strongly suppressed.  In addition, the Wino LSP and its charged partners may be observable at colliders.  Furthermore, flavor violation arising from SUSY particle exchange such as $\mu \rightarrow e \gamma$ are highly suppressed in this scenario, owing to the large masses of the sleptons and squarks.    

\section{Discussion and conclusion}

In this paper we have resurrected the minimal Yukawa sector of SUSY $SO(10)$, which explains the entirety of fermion masses and mixings in terms of two symmetric Yukawa matrices.  While the fermion masses and mixing angles were known to fit well with such a structure arising from a $10_H$ and a
$\overline{126}_H$ of Higgs fields, once they are coupled minimally to a $\{126_H + 210_H\}$ Higgs sector, the model becomes inconsistent. This is because of the tight nature of symmetry breaking and fermion mass generation. The scale of right-handed neutrino masses is required from light neutrino mass spectrum to be  $V_R \sim 10^{12}$ GeV, while for such values of $V_R$, the masses of several colored chiral superfields become of order
$V_R^2/M_{GUT} \sim 10^{10}$ GeV.  This causes the gauge couplings to become non-perturbative before reaching the GUT scale, causing inconsistencies.

Our proposal to fix this problem is to introduce a $54_H$ that has no Yukawa couplings to the SM fermions.  This enables a reduction of the $B-L$ symmetry breaking scale to $V_R \sim 10^{12}$ GeV, without causing any of the chiral superfields to remain light (except for a pair of SM singlets needed for $B-L$ symmetry breaking).  The supersymmetric sector of such a theory has a relatively small number of parameters, and the predictions arising from the Yukawa sector now become reliable and consistent.

We have performed a fermion fit in this scenario including threshold effects from $V_R$ to $M_{GUT}$.  In addition, proton decay constraints require the SUSY spectrum to be mini-split, with the gauginos and Higgsinos having masses of order TeV and the squarks and sleptons of order 100 TeV.  In our new fit to fermion data we have incorporated the mini-split SUSY spectrum as well. Such a scenario has a natural embedding in pure gravity mediation.  In this rendition proton lifetime for decay into $\overline{\nu} K^+$ cannot exceed another order of magnitude compared to the present lower limit of $5.9 \times 10^{33}$ yrs.  The model can also be tested at colliders with the discovery of Wino LSP along with its charged partners.

\section*{Acknowledgments}
This work is supported in part by the U.S. Department of Energy Grant No.  de-sc0016013  (K.S.B and S.S). BB acknowledges the financial support from the Slovenian Research Agency (research core funding No.~P1-0035) and
thanks the Department of Physics of the Oklahoma State University for hospitality. The work of K.S.B is supported in part by a Fermilab Distinguished Scholar program. K.S.B and S.S  would like to thank the Fermilab theory group for
the hospitality during the 2017 summer visit where part of the work was done. We have benefited from discussions with  T. Fukuyama, S. Khan and R. Mohapatra. K.S.B and BB acknowledge the hospitality of  INT, Seattle during 2017  Neutron-Antineutron Oscillations workshop.

\begin{appendices}
\appendixpageoff

\section{Details of proton decay calculation}\label{B}
Here we provide more details on the proton decay calculation.

Corresponding to the solution of the stationary conditions Eqs. \eqref{SE-1} -\eqref{SE-2}, the $4 \times 4$ Higgs doublet mass matrix is given by
\begin{align}\label{doublet}
M_D=
\left(
\begin{array}{cccc}
 m_3+\sqrt{\frac{3}{5}} V_E \lambda _{13} &
   \frac{\left(\sqrt{2} V_2-V_3\right)
   \lambda _3}{2 \sqrt{5}} &
   -\frac{\left(\sqrt{2} V_2+V_3\right)
   \lambda _4}{2 \sqrt{5}}&-\frac{\lambda_4 V_R}{\sqrt{5}} \\
 \frac{\left(\sqrt{2} V_2-V_3\right) \lambda
   _4}{2 \sqrt{5}} & -\frac{\left(2 \sqrt{2}
   V_2^2+16 V_3 V_2+\sqrt{2} V_3^2\right)
   \lambda _2}{120 V_2} & \frac{\left(V_3^2-2
   V_2^2\right) \lambda _1 \lambda _{12}}{15
   \sqrt{2} V_2 \lambda _{10}} &0\\
 -\frac{\left(\sqrt{2} V_2+V_3\right) \lambda
   _3}{2 \sqrt{5}} & \frac{\left(V_3^2-2
   V_2^2\right) \lambda _1 \lambda _{11}}{15
   \sqrt{2} V_2 \lambda _{10}} &
   -\frac{\left(2 \sqrt{2} V_2^2+8 V_3
   V_2+\sqrt{2} V_3^2\right) \lambda _2}{120
   V_2} & \frac{\lambda_2 V_R}{10}\\
   -\frac{\lambda_3 V_R}{\sqrt{5}}&0&\frac{\lambda_2 V_R}{10}&m_{44}
\end{array}
\right),
\end{align}

\noindent
where
\begin{align}
m_{44}= \frac{\lambda_1}{10 V_2} \left(  2 \sqrt{2} V^2_2 + 5 V_2 V_3 - \sqrt{2} V^2_3  \right) -\frac{1}{4} \sqrt{\frac{3}{5}} \lambda_{10} V_E.
\end{align}
The matrix in Eq. (\ref{doublet}) is written in a basis where the row vector

$$\{H_{(1,2,2)}^{(1, 2,-1/2)}, \, \overline{\Delta}_{(15,2,2)}^{(1,
2,-1/2)},\, \Delta_{(15,2,2)}^{(1, 2,-1/2)},\,
\Phi_{(6,2,2)}^{(1, 2,-1/2)}\}$$
multiplies the matrix from the left and the column vector
$$\{H_{(1,2,2)}^{(1, 2,1/2)}, \,\Delta_{(15,2,2)}^{(1,2,1/2)},\, \overline{\Delta}_{(15,2,2)}^{(1, 2,1/2)},\,
\Phi_{(6,2,2)}^{(1, 2,1/2)}\}^T$$
 multiplies from the right. Here we have used the decomposition under $SU(4)_c \times SU(2)_L \times SU(2)_R$ as subscripts and the decomposition under SM as superscripts.

The determinant of the Higgs doublet mass matrix should be near zero.  This condition is achieved by a fine-tuning, which
we use to determine $m_3$. Note that, in the limit $V_R=0$, the Higgs doublet from the $210_H$ decouples from the rest, hence only the upper $3\times 3$ block is relevant to an excellent approximation.
The MSSM Higgs superfields $H_u$ and $H_d$ are mixtures of the doublets coming from $10_H, 126_H$ and $\overline{126}_H$ multiplets denoted as
\begin{align}
&H_u=N_u (H_u^{10}+ p_2 H_u^{126}+ p_3 H_u^{\overline{126}}),\\
&H_d=N_d (H_d^{10}+ q_2 H_d^{\overline{126}}+ q_3 H_d^{126}).
\end{align}

\noindent The normalization factors are defined as $N_u= \frac{1}{\sqrt{1+|p_2|^2+|p_3|^2}}$ and $N_d= \frac{1}{\sqrt{1+|q_2|^2+|q_3|^2}}$. The  expressions for $p_i$ and $q_i$ can be found in a straightforward way from the left and the right eigenvectors corresponding to the zero eigenvalue of Eq. (\ref{doublet}):
\begin{align}
&p_2^{\ast}=
\scalemath{0.8}{
\frac{6 \sqrt{5} \lambda _{10} V_2 \left(\lambda _2
   \lambda _4 \lambda _{10} \left(4 V_2^3+6
   \sqrt{2} V_3 V_2^2-6 V_3^2 V_2-\sqrt{2}
   V_3^3\right)+4 \lambda _1 \lambda _3
   \lambda _{12} \left(2 V_2+\sqrt{2}
   V_3\right) \left(2
   V_2^2-V_3^2\right)\right)}{\lambda _2^2 \lambda _{10}^2 \left(4 V_2^4+24
   \sqrt{2} V_3 V_2^3+68 V_3^2 V_2^2+12
   \sqrt{2} V_3^3 V_2+V_3^4\right)-16 \lambda
   _1^2 \lambda _{11} \lambda _{12}
   \left(V_3^2-2 V_2^2\right){}^2}
   },
\\
&p_3^{\ast}=
\scalemath{0.8}{
-\frac{6 \sqrt{10} \lambda _{10} V_2 \left(\lambda _2
   \lambda _3 \lambda _{10} \left(2 \sqrt{2}
   V_2^3+18 V_3 V_2^2+9 \sqrt{2} V_3^2
   V_2+V_3^3\right)+4 \lambda _1 \lambda _4
   \lambda _{11} \left(\sqrt{2} V_2-V_3\right)
   \left(2 V_2^2-V_3^2\right)\right)}{\lambda _2^2 \lambda _{10}^2 \left(4 V_2^4+24
   \sqrt{2} V_3 V_2^3+68 V_3^2 V_2^2+12
   \sqrt{2} V_3^3 V_2+V_3^4\right)-16 \lambda
   _1^2 \lambda _{11} \lambda _{12}
   \left(V_3^2-2 V_2^2\right){}^2}
} ,
\end{align}

\begin{align}
&q_2^{\ast}=
\scalemath{0.8}{
\frac{6 \sqrt{5} \lambda _{10} V_2 \left(\lambda _2
   \lambda _3 \lambda _{10} \left(4 V_2^3+6
   \sqrt{2} V_3 V_2^2-6 V_3^2 V_2-\sqrt{2}
   V_3^3\right)+4 \lambda _1 \lambda _4
   \lambda _{11} \left(2 V_2+\sqrt{2}
   V_3\right) \left(2
   V_2^2-V_3^2\right)\right)}{\lambda _2^2 \lambda _{10}^2 \left(4 V_2^4+24
   \sqrt{2} V_3 V_2^3+68 V_3^2 V_2^2+12
   \sqrt{2} V_3^3 V_2+V_3^4\right)-16 \lambda
   _1^2 \lambda _{11} \lambda _{12}
   \left(V_3^2-2 V_2^2\right){}^2}
}, \label{q2}
\\
&q_3^{\ast}=
\scalemath{0.8}{
-\frac{6 \sqrt{10} \lambda _{10} V_2 \left(\lambda _2
   \lambda _4 \lambda _{10} \left(2 \sqrt{2}
   V_2^3+18 V_3 V_2^2+9 \sqrt{2} V_3^2
   V_2+V_3^3\right)+4 \lambda _1 \lambda _3
   \lambda _{12} \left(\sqrt{2} V_2-V_3\right)
   \left(2 V_2^2-V_3^2\right)\right)}{\lambda _2^2 \lambda _{10}^2 \left(4 V_2^4+24
   \sqrt{2} V_3 V_2^3+68 V_3^2 V_2^2+12
   \sqrt{2} V_3^3 V_2+V_3^4\right)-16 \lambda
   _1^2 \lambda _{11} \lambda _{12}
   \left(V_3^2-2 V_2^2\right){}^2}
}.
\end{align}

\noindent With these definitions,  the parameters $r$ and $s$ of Eq. \eqref{rs-original} which are determined from the fermion mass fit can be expressed in terms of the superpotential parameters:
\begin{align}\label{rs}
r=\frac{N_u}{N_d}, \;\;\;
 s^{\ast}=\frac{p_3}{q_2}.
\end{align}
In our numerical scan for proton decay amplitude, we impose the constraints on $r$ and $s$ from fermion mass fit in conjunction with Eq. (\ref{rs}).

The mass matrix for the color triplets Higgs(inos) mediating $d=5$ proton decay is given by
\begin{align}\label{triplet}
M_T=
\left(
\scalemath{0.85}{
\begin{array}{ccccc}
 m_3- \frac{2 V_E \lambda _{13}}{\sqrt{15}} &
   \frac{\left(2 V_2^2-V_3^2\right) \lambda
   _3}{2 \sqrt{30} V_2} & -\frac{\left(2
   V_2^2+V_3^2\right) \lambda _4}{2 \sqrt{30}
   V_2} & -\sqrt{\frac{2}{15}} V_3 \lambda _4&\frac{\lambda_4 V_R}{\sqrt{5}}
   \\
 \frac{\left(2 V_2^2-V_3^2\right) \lambda
   _4}{2 \sqrt{30} V_2} & -\frac{\left(6
   \sqrt{2} V_2^2+12 V_3 V_2+\sqrt{2}
   V_3^2\right) \lambda _2}{120 V_2} &
   \frac{\sqrt{2} \left(V_3^2-2 V_2^2\right)
   \lambda _1 \lambda _{12}}{15 V_2 \lambda
   _{10}} & 0 &0\\
 -\frac{\left(2 V_2^2+V_3^2\right) \lambda
   _3}{2 \sqrt{30} V_2} & \frac{\sqrt{2}
   \left(V_3^2-2 V_2^2\right) \lambda _1
   \lambda _{11}}{15 V_2 \lambda _{10}} &
   -\frac{\left(6 \sqrt{2} V_2^2+12 V_3
   V_2+\sqrt{2} V_3^2\right) \lambda _2}{120
   V_2} & \frac{V_3 \lambda _2}{15 \sqrt{2}}&-\frac{\lambda_2 V_R}{10 \sqrt{3}}
   \\
 -\sqrt{\frac{2}{15}} V_3 \lambda _3 & 0 &
   \frac{V_3 \lambda _2}{15 \sqrt{2}} &
   -\frac{1}{30} \left(\sqrt{2} V_2+3
   V_3\right) \lambda _2 &-\frac{\lambda_2 V_R}{5 \sqrt{6}}
   \\
 \frac{\lambda_3 V_R}{\sqrt{5}}&0&-\frac{\lambda_2V_R}{10 \sqrt{3}}&-\frac{\lambda_2 V_R}{5 \sqrt{6}}&m_{55}
\end{array}
}
\right),
\end{align}

\noindent where
\begin{align}
m_{55}= -\frac{\lambda_1}{60 V_2} \left( 8 \sqrt{2} V^2_2 - 40 V_2 V_3 + \sqrt{2} V^2_3  \right) + \frac{\lambda_{10}}{2 \sqrt{15}} V_E.
\end{align}
The matrix in Eq. (\ref{triplet}) is written in a basis where the row vector\newline
$$\{H_{(6,1,1)}^{(\overline{3}, 1,1/3)}, \, \overline{\Delta}_{(6,1,1)}^{(\overline{3},
1,1/3)},\, \Delta_{(6,1,1)}^{(\overline{3}, 1,1/3)},\, \Delta_{(\overline{10},1,3)}^{(\overline{3}, 1,1/3)},\,
\Phi_{(15,1,3)}^{(\overline{3}, 1,1/3)}\}$$
multiplies the matrix from the left and the column vector
$$\{H_{(6,1,1)}^{(3, 1,-1/3)}, \,\Delta_{(6,1,1)}^{(3,1,-1/3)},\, \overline{\Delta}_{(6,1,1)}^{(3, 1,-1/3)},\,\overline{\Delta}_{(10,1,3)}^{(3, 1,-1/3)},\,
\Phi_{(15,1,3)}^{(3, 1,-1/3)}\}^T$$
 multiplies from the right.

In the limit $V_R=0$,  only the upper $4\times 4$ block of Eq. (\ref{triplet}) is relevant. We thus work in the approximation that $210_H$ contribution decouples. Note that the decoupled color triplets from $210_H$ do not lead to $d=5$ proton decay.

The relevant  nuclear matrix elements for proton decay calculation in our model are \cite{Aoki:2017puj}:
\begin{align}
\langle K^+|(us)_Ld_L|p\rangle &= 0.041 \;\;\textrm{GeV}^2,
\\
 \langle K^+|(ud)_Ls_L|p\rangle &= 0.139 \;\;\textrm{GeV}^2,\\
 \langle \pi^+|(ud)_Ld_L|p\rangle &= 0.189 \;\;\textrm{GeV}^2. \label{nuc}
\end{align}

\noindent 
We use as input $m_p=0.94$ GeV, $m_K=0.494$ GeV,  $m_{\pi}=0.139$ GeV, $m_{wino}=1$ TeV and $g_2(m_{susy})=0.6518$. For numerical calculations, we also take $R_L=1.25$, and $A_S=6.54$ from Ref. \cite{Babu:2012pb}, where $R_L$ ans $A_S$ are the long
distance and short distance renormalization factors of the corresponding $LLLL$ d=5 operators respectively.


\section{RGEs from TeV scale to SUSY scale}\label{C}

In this Appendix we collect the relevant RGEs for split-SUSY relevant for evolution from 1 TeV to the squark and slepton mass scale ($\tilde{m}$) of order 100 TeV.   The 2-loop renormalization-group equations for the gauge couplings are given by \cite{Giudice:2004tc}:
\bea
(4\pi)^2\frac{d}{dt}~ g_i=g_i^3b_i &+&\frac{g_i^3}{(4\pi)^2}
\left[ \sum_{j=1}^3B_{ij}g_j^2-\sum_{\alpha=u,d,e}d_i^\alpha
{\rm Tr}\left( y^{\alpha \dagger}y^{\alpha}\right)\right. \nonumber \\
&-&\left. d_W\left(
\tilde{g}^2_u +\tilde{g}^2_d \right)
-d_B\left( \tilde{g}^{\prime 2}_u +\tilde{g}^{\prime 2}_d \right)\right] ,
\label{gauger}
\eea
where $t=\ln \bar \mu$ with $\bar \mu$ being the renormalization scale.
Here the convention used is $g_1^2=(5/3)g^{\prime 2}$.
Eq. \eqref{gauger} is scheme-independent
up to the two-loop order.

In the effective theory below $\tilde{m}$, the $\beta$-function coefficients
are
\bea
&&b=\left(\frac{9}{2},-\frac{7}{6},-5\right), \;\;\; B= \begin{pmatrix}
\frac{104}{25}&
\frac{18}{5}&\frac{44}{5}\cr \frac{6}{5} & \frac{106}{3}&12 \cr
\frac{11}{10}&\frac{9}{2}&22
\end{pmatrix}, \\
&&d^u=\left(\frac{17}{10},\frac{3}{2},2\right), \;\;\;
d^d=\left(\frac{1}{2},\frac{3}{2},2\right), \;\;\;
d^e=\left(\frac{3}{2},\frac{1}{2},0\right) \\
&&d^W=\left(\frac{9}{20},\frac{11}{4},0\right), \;\;\;
d^B=\left(\frac{3}{20},\frac{1}{4},0\right) .
\eea

For the Yukawa coupling evolution to one loop order, one has
\bea
(4\pi)^2\frac{d}{dt}~y^u&=&y^u\left( -3\sum_{i=1}^3c_i^ug_i^2+\frac{3}{2}y^{u \dagger}y^{u}-\frac{3}{2}y^{d \dagger}y^{d}+T\right),\\
(4\pi)^2\frac{d}{dt}~y^d&=&y^d\left( -3\sum_{i=1}^3c_i^dg_i^2
-\frac{3}{2}y^{u \dagger}y^{u}+\frac{3}{2}y^{d \dagger}y^{d}+T\right),\\
(4\pi)^2\frac{d}{dt}~y^e&=&y^e\left( -3\sum_{i=1}^3c_i^eg_i^2+\frac{3}{2}y^{e \dagger}y^{e}+T\right),
\label{yukk}
\eea
where,
\beq
T={\rm Tr} \left( 3y^{u \dagger}y^{u}+3 y^{d \dagger}y^{d}+y^{e \dagger}y^{e}\right) +\frac{3}{2}\left( \tilde{g}^2_u +\tilde{g}^2_d \right)+\frac{1}{2}\left( \tilde{g}^{\prime 2}_u +\tilde{g}^{\prime 2}_d \right)
\eeq
\beq
c^u=\left( \frac{17}{60}, \frac{3}{4}, \frac{8}{3}\right), \;\;\;
c^d=\left( \frac{1}{12}, \frac{3}{4}, \frac{8}{3}\right), \;\;\;
c^e=\left( \frac{3}{4}, \frac{3}{4}, 0\right) .
\eeq

The renormalization-group equations for the gaugino couplings  are:
\bea
(4\pi)^2\frac{d}{dt}~\tilde{g}_u &=&-3\tilde{g}_u \sum_{i=1}^3 C_i g_i^2
+\frac{5}{4}\tilde{g}_u^3 -\frac{1}{2} \tilde{g}_u \tilde{g}_d^2
+\frac{1}{4} \tilde{g}_u \tilde{g}^{\prime 2}_u
+\tilde{g}_d \tilde{g}^{\prime}_d \tilde{g}^{\prime}_u +\tilde{g}_u T \\
(4\pi)^2\frac{d}{dt}~\tilde{g}_u^{\prime} &=&-3\tilde{g}_u^{\prime} \sum_{i=1}^3 C_i^{\prime} g_i^2
+\frac{3}{4}\tilde{g}_u^{\prime 3} +\frac{3}{2} \tilde{g}_u^{\prime} \tilde{g}_d^{\prime 2}
+\frac{3}{4} \tilde{g}_u^{\prime} \tilde{g}_u^2
+3\tilde{g}_d^{\prime} \tilde{g}_d \tilde{g}_u + \tilde{g}_u^{\prime} T \\
(4\pi)^2\frac{d}{dt}~\tilde{g}_d &=&-3\tilde{g}_d \sum_{i=1}^3 C_i g_i^2
+\frac{5}{4}\tilde{g}_d^3 -\frac{1}{2} \tilde{g}_d \tilde{g}_u^2
+\frac{1}{4} \tilde{g}_d \tilde{g}_d^{\prime 2}
+\tilde{g}_u \tilde{g}_u^{\prime} \tilde{g}_d^{\prime} +\tilde{g}_d T \\
(4\pi)^2\frac{d}{dt}~\tilde{g}_d^{\prime} &=&-3\tilde{g}_d^{\prime} \sum_{i=1}^3 C_i^{\prime} g_i^2
+\frac{3}{4}\tilde{g}_d^{\prime 3} +\frac{3}{2} \tilde{g}_d^{\prime} \tilde{g}_u^{\prime 2}
+\frac{3}{4} \tilde{g}_d^{\prime} \tilde{g}_d^2
+3\tilde{g}_u^{\prime} \tilde{g}_u \tilde{g}_d +\tilde{g}_d^{\prime} T ,
\eea
\beq
{\rm with} ~~~~ C=\left( \frac{3}{20},\frac{11}{4},0\right),\;\;\;
C^\prime =\left( \frac{3}{20},\frac{3}{4},0\right) .
\eeq

 The one-loop RGE  for the Higgs quartic coupling (which appears in the RGE for the $d=5$ neutrino mass operator) is given by
\bea
(4\pi)^2\frac{d}{dt}~\lambda  &=&12 \lambda^2 +\lambda \left[
-9 \left( \frac{g_1^2}{5}+g_2^2\right)+6\left( \tilde{g}_u^2 +\tilde{g}_d^2 \right)
+2\left( \tilde{g}_u^{\prime 2} +\tilde{g}_d^{\prime 2} \right) \right. \nonumber \\
&&\left. +4 {\rm Tr}(3y^{u \dagger}y^{u}+
3y^{d \dagger}y^{d}
+y^{e \dagger}y^{e})\right] +\frac 92 \left( \frac{g_2^4}{2}+
\frac{3g_1^4}{50}+\frac{g_1^2g_2^2}{5}
\right) \nonumber \\
&& -5\left( \tilde{g}_u^4 +\tilde{g}_d^4 \right) -2 \tilde{g}_u^2\tilde{g}_d^2
-\left( \tilde{g}_u^{\prime 2} +\tilde{g}_d^{\prime 2} \right)^2  -2
\left( \tilde{g}_u \tilde{g}_u^{\prime} + \tilde{g}_d \tilde{g}_d^{\prime} \right)^2 \nonumber \\
&&-4 {\rm Tr}\left[3(y^{u \dagger}y^{u})^2+
3(y^{d \dagger}y^{d})^2
+(y^{e \dagger}y^{e})^2 \right] .
\eea

We have extended the one-loop RGE for the $d=5$ neutrino mass operator to the case of split-SUSY:
\begin{eqnarray} \label{eq:finalrge}
        16\pi^2 \beta_\kappa & = &
         -\frac{3}{2} \left[\kappa \left( Y_e^\dagger Y_e \right)
         +            \left( Y_e^\dagger Y_e \right)^T \kappa \right]
         +\lambda \kappa - 3 g_2^2 \kappa
\nonumber\\
         & &
         +2 \, \left[ \Tr \left( 3 Y_u^\dagger Y_u + 3 Y_d^\dagger Y_d
         +Y_e^\dagger Y_e \right) + \frac{3}{2} \left( \tilde{g}^2_u + \tilde{g}^2_d \right) + \frac{1}{2} \left( \tilde{g}^{\prime 2}_u + \tilde{g}^{\prime 2}_d \right)  \right] \kappa.
\end{eqnarray}

\noindent The neutrino mass matrix is given by $\mathcal{M}_{\nu}= -\frac{v^2}{2} \kappa$, with $v \simeq 174$ GeV.

\section{RGEs from $V_R$ to $M_{GUT}$}\label{D}

Between $V_R$ and $M_{GUT}$ we have  $SM \times U(1)_{B-L}$. Due to the presence of the extra $U(1)$ gauge boson, the RGEs of the Yukawa couplings get modified and are relevant for our study. Furthermore, the right-handed neutrinos have masses of order $V_R$, and they will contribute to the evolution of parameters above $V_R$. Here we provide the new RGEs including these effects.  The coupling of $\nu^c$ with a SM singlet field $\overline{\sigma}$ is defined to be $W \supset \frac{Y_R}{2} \nu^c \nu^c \overline{\sigma}$.
\begin{align}
&16 \pi^2 \frac{d Y_D}{dt} = Y_D \left[  3 Y^{\dagger}_D Y_D +  Y^{\dagger}_U Y_U + Tr \left( Y^{\dagger}_E Y_E \right) + 3 Tr \left( Y^{\dagger}_D Y_D \right) -\frac{7}{15} g^2_1 -3 g^2_2 -\frac{16}{3} g^2_3 - \frac{1}{6} g^2_B  \right] \label{th1} \\
&16 \pi^2 \frac{d Y_U}{dt}= Y_U \left[  3 Y^{\dagger}_U Y_U +  Y^{\dagger}_D Y_D + Tr \left( Y^{\dagger}_{\nu_D} Y_{\nu_D} \right) + 3 Tr \left( Y^{\dagger}_U Y_U \right) -\frac{13}{15} g^2_1 -3 g^2_2 -\frac{16}{3} g^2_3 - \frac{1}{6} g^2_B  \right]\\
&16 \pi^2 \frac{d Y_E}{dt}= Y_E \left[  3 Y^{\dagger}_E Y_E + Y^{\dagger}_{\nu_D} Y_{\nu_D} + 3 Tr \left( Y^{\dagger}_D Y_D \right) -\frac{9}{5} g^2_1 -3 g^2_2 - \frac{3}{2} g^2_B  \right]\\
&16 \pi^2 \frac{dY_{\nu_D}}{dt}= Y_{\nu_D} \left[  3 Y^{\dagger}_{\nu_D} Y_{\nu_D} + Y^{\dagger}_E Y_E + Tr \left( Y^{\dagger}_{\nu_D} Y_{\nu_D} \right) + 3 Tr \left( Y^{\dagger}_U Y_U \right) + Y^{\dagger}_R Y_R -\frac{3}{5} g^2_1 -3 g^2_2 - \frac{3}{2} g^2_B  \right] \\
&16 \pi^2 \frac{dY_R}{dt}= Y_R \left( Y^{\dagger}_{\nu_D} Y_{\nu_D} + Y^{\dagger}_R Y_R  \right) + \left( Y^{T}_{\nu_D} Y^{\ast}_{\nu_D} + Y^{\ast}_R Y_R  \right) Y_R + Y_R \left[ Tr \left( Y^{\dagger}_R Y_R \right) -\frac{9}{4} g^2_B \right] , \label{th2}
\end{align}

\noindent where $g_B$ is the $B-L$ gauge coupling and $Y_R(M_{GUT}) =Y_{126}(M_{GUT})$.  The $SU(3)_c$ and the $SU(2)_L$ gauge couplings evolve as in MSSM, but due to the presence of two $U(1)$ factors, the corresponding gauge bosons mix kinetically \cite{Holdom:1985ag,delAguila:1988jz}. The renormalization group equations for the coupling-constant flow of $U(1)_Y\times U(1)_{B-L}$ are given by \cite{Babu:1996vt} (see also \cite{Fonseca:2011vn,Fonseca:2013bua}): 
\begin{align}
&16 \pi^2 dg_Y/dt=  g_Y^3 B_{YY}\\
&16 \pi^2 dg_B/dt=  g_B \left(  g_B^2 B_{BB} + g_{YB}^2 B_{YY} + 2 g_B g_{YB} B_{YB} \right) \\
&16 \pi^2 dg_{YB}/dt=   g^3_{YB} B_{YY}+g^2_{B}g_{YB} B_{BB}+2 g^2_{Y}g_{YB} B_{YY}+2 g^2_{Y}g_B B_{YB}+2 g_B g^2_{YB} B_{YB}  
\end{align}
\noindent with $B_Y=33/5, B_B=9, B_{YB}=6\sqrt{2/5}$. Even though $g_{YB}$ is zero at the GUT scale due to the kinetic mixing, $g_{YB}$ of order 0.04 is induced in going from GUT scale to $V_R$. This changes the value of $g_B$ at $V_R$ by about $5\%$ compared to the case of ignoring $B_{YB}$. This can be safely ignored for the analysis performed in this work, however, it can be important for  precise gauge coupling unification.

\end{appendices}

\end{document}